\documentclass[twocolumn,showpacs,prb,amsmath,amssymb,floatfix]{revtex4-1}

\usepackage{placeins}
\usepackage{epsfig}
\usepackage{amsmath}
\usepackage{amsfonts}
\usepackage{amssymb}
\usepackage{amsbsy}
\usepackage{mathrsfs}
\usepackage{graphicx}
\usepackage{subfigure}
\usepackage{dcolumn}
\usepackage{bm}

\newcommand{\rr}{\boldsymbol{r}}
\newcommand{\kk}{\boldsymbol{k}}
\newcommand{\va}{\boldsymbol{a}}
\newcommand{\vb}{\boldsymbol{b}}
\newcommand{\vc}{\boldsymbol{c}}
\newcommand{\vd}{\boldsymbol{d}}

\renewcommand{\v}[1]{\boldsymbol{#1}}

\renewcommand{\d}{\hspace{.5mm}\mathrm{d}\hspace{.5mm}}
\newcommand{\dd}{\hspace{.5mm}\mathrm{d}^2}

\DeclareMathOperator{\sinc}{sinc}
\DeclareMathOperator{\sign}{sign}

\newcommand{\cre}{\hat{a}^{\dagger}}
\newcommand{\ann}{\hat{a}}

\newcommand{\braket}[3]{\left\langle #1\,\left|#2\right|#3\right\rangle}

\begin{document}

\title{Lagrange mesh and exact diagonalization for numerical study of semiconductor quantum dot systems with application in singlet-triplet qubits}

\author{Tuukka Hiltunen}
\author{Juha Ritala}
\author{Oona Kupiainen} 
\author{Topi Siro}
\author{Ari Harju}

\affiliation{ 
  Department of Applied Physics and Helsinki Institute of Physics,
  Aalto University School of Science, P.O. Box 14100, 00076 Aalto, Finland
}

\date{\today}

\begin{abstract}

We present a highly flexible computational scheme for studying correlated electrons confined by 
an arbitrary external potential in two-dimensional semiconductor quantum dots.
The method starts by a Lagrange mesh calculation for the single-particle states, followed 
by the calculation of the Coulomb interaction matrix elements between these, and 
combining both in the exact diagonalization of the many-body Hamiltonian.
We apply the method in simulation of double quantum dot singlet-triplet qubits.
We simulate the full quantum control and dynamics of one singlet-triplet qubit. We also use our method to provide
an exact diagonalization based first-principles model for studying two singlet-triplet qubits and their capacitative coupling
via the long-distance Coulomb interaction.
\end{abstract}

\maketitle

\section{Introduction}

The development of experimental methods
has enabled the fabrication of ``artificial atoms" 
with a controlled
number of electrons, ranging from a few to a few hundred, confined in
a tunable external potential inside a semiconductor.\cite{Ashoori96,Reimann02,Saarikoski_RMP}
These quantum dots (QD's) have been
proposed as a possible realization for the qubit of a quantum
computer \cite{Loss98,Hanson07}.

A framework for using two-electron spin eigenstates as qubits was proposed by Levy in 2002 \cite{levy}.
The two-electron double quantum dot (DQD) spin states have natural protection against the decoherence by the hyperfine
interaction and allow a scalable architecture for quantum computation\cite{taylor}.
The universal set of quantum gates for two spin singlet-triplet DQD qubits has been demonstrated
experimentally. These gates include one qubit rotations
generated by the exchange interaction\cite{Pettaa} and stabilized hyperfine magnetic field
gradients\cite{foletti}, and two qubit operations using long distance capacitative coupling by the Coulomb\cite{shulman} interaction.

In creating the inter-qubit gates and operations, quantum entanglement is essential\cite{ent}.
The aforementioned capacitative coupling is one possible method to create entangled states and
implement two-qubit operations in singlet-triplet qubits, the other possibility being exchange based methods\cite{levy,li}.
In the capacitative dipole-dipole coupling, the entanglement is achieved by differing charge densities in the singlet and triplet states
that result in different Coulomb repulsion between the qubits. This conditioning can be used to create entangled states and to implement
the two-qubit gates required for universal quantum computing\cite{taylor,hanson,shulman,stepa,Nielsen}.

Although other methods, like the variational quantum Monte Carlo\cite{VMC} and the density functional
theory\cite{Henri}, have shown to give reasonably accurate results, exact diagonalization
is still the most reliable technique for small particle numbers.
In this paper, we use the Lagrange mesh method\cite{Baye86} and exact diagonalization to
simulate one and two singlet-triplet qubit systems. The one-particle eigenstates are computed using the 
Lagrange mesh method and are then used to create the many-body basis
in the calculations.
The use of the Lagrange mesh method as a source of single-particle states
allows us to study very flexibly various forms of the confinement potential. For example,
the relatively complex two-DQD case can be handled with ease using this method.
Still, the one-body basis needed for good accuracy is much more compact in the Lagrange mesh than, 
e.g., in the finite-difference formulation.

This paper is organized as follows. In Section II, the theoretical model used in our computations is briefly discussed.
In Section III, we introduce the Lagrange mesh method for many-body problems in quantum dots.
The computational results are shown in Section IV. The single-particle states and their convergence are discussed in
Section IV A. In Section IV B, we model the full quantum control and dynamics of a singlet-triplet qubit. In Section IV C,
we use the Lagrange mesh method to create a realistic first-principles ED model for studying the interplay and entanglement of two
singlet-triplet qubits.

\section{Model}

We model a lateral GaAs quantum dot system with the two-dimensional Hamiltonian
\begin{equation}\nonumber%
\label{Hamiltonian}
\hat{H}=\sum^N_{j=1}\left[\frac{({\v{p}}_j+e\v A(\rr_j)
      )^2} {2 m^*}+V(\rr_j)+V_Z(\rr_j)\right] +\sum_{j<k}\frac{e^2}{4\pi\epsilon r_{jk}} ,
\end{equation}
where $V_Z(\rr_j)=g^*\mu_B\v{B}(\v{r}_j)\cdot\v{S}_j$ is the Zeeman term with the effective GaAs g-factor
$g^*=-0.44$. $\v{A}$ is the magnetic vector potential, and $m^*\approx0.067\,m_e$ and
$\epsilon\approx12.7\,\epsilon_0$ are the effective electron mass and
permittivity in GaAs, respectively. In numerical work, it is
convenient to switch into effective atomic units by setting
$m^*=e=\hbar=1/4\pi \epsilon=1$. In these units, energy is given by
$\mathrm{Ha}^* \approx 11.30$~meV and length in $a_0^* \approx 10.03$~nm.

In our computations, the external potential $V(\v{r})$ for quantum dot systems
consists of several parabolic wells. A confinement potential of $n$ parabolic wells can be written as
\begin{equation}\label{eq:qdpot}
V(\v{r})=\frac{1}{2}m^*\omega_0^2\min_{1\leq j \leq n}\{|\v{r}-\v{r}_j|^2\},
\end{equation}
where $\{\v{r}_j\}_{1\leq j\leq n}$ are the locations of the minima of the parabolic wells, and $\omega_0$ is the
confinement strength. The kinks caused by the $\min$-function are smoothed in the potential.

\section{Method}

The Lagrange mesh method\cite{Baye86} is a very efficient method
for solving the Schr\"{o}dinger equation, and it has the simplicity of a
finite-difference mesh calculation, since no integrations need to be performed.
It also does not suffer from the same limitations regarding to the confinement potential
as for example using the analytical Fock-Darwin basis.

We will use this technique to solve the eigenstates of the one particle Hamiltonian,
\begin{equation}\label{eq:h1}
\hat{H}=\frac{\v{p}^2
      } {2 m^*}+V(\rr),
\end{equation}
omitting the magnetic vector potential here.
The eigenstates are then used as a basis for the many-body calculation done by
the exact diagonalization technique, and the Zeeman term and Coulomb interaction are included in it.

\subsection{One-particle problem}

A set of $N$ Lagrange functions $L_k$ defined over an interval $(a,b)$ is associated with $N$ mesh points $x_k\in(a,b)$ and a corresponding Gauss quadrature
\begin{equation}\label{GaussQuadrature}
 \int_{a}^{b}\d x f(x) \approx \sum_{k=1}^N\lambda_k f(x_k)\,.
\end{equation}
The Lagrange functions are infinitely differentiable real functions, which are orthonormal,
\begin{equation}\label{orthonormality}
 \int_a^b \d xL_i(x)L_j(x) =\delta_{ij}\,,
\end{equation} 
and satisfy the Lagrange conditions
\begin{equation}\label{LagrangeCondition}
 L_{i}(x_j)=\lambda_i^{-1/2}\delta_{ij}\,.
\end{equation}
From the conditions of Eqs.~\eqref{orthonormality} and \eqref{LagrangeCondition}, it follows that the Gauss quadrature is exact for any product of two Lagrange functions:
\begin{equation}\nonumber
 \int_a^b \d xL_i(x)L_j(x) =\delta_{ij}=\sum_{k=1}^N\lambda_kL_{i}(x_k)L_{j}(x_k)\,.
\end{equation} 

Many different Lagrange meshes, mostly based on orthogonal polynomials
or trigonometric functions, have been proposed \cite{Baye06} both for
finite intervals and the infinite intervals $(0,\infty)$ and
$(-\infty,\infty)$. The meshes can also be modified to distribute the
mesh points optimally for a particular system.\cite{Varga04}

One of the most simple Lagrange meshes is the sinc mesh.\cite{Baye06} It is
defined over the interval $(-\infty,\infty)$, but designed to treat fairly
well localized wave functions. The mesh points distributed uniformly around
the origin are
\begin{equation}\label{MeshPoints}
 x_a=a\,, \ \  a\in \left\{-\tfrac{N-1}{2}\,,\,-\tfrac{N-1}{2}+1\,,\,\ldots\,,\,\tfrac{N-1}{2}\right\}\,,
\end{equation} 
and all the weights in the Gauss quadrature are $\lambda_a=1$. The Lagrange-sinc functions are

\begin{equation}\nonumber
 L_a(x)=\sinc(x-a)=\frac{\sin[\pi(x-a)]}{\pi(x-a)}\,.
\end{equation} 
The matrix elements of the derivatives $\partial_x$ and $\partial_x^2$
between two sinc functions can be calculated analytically, resulting
in
\begin{eqnarray}\nonumber
 \left(\partial_x\right)_{a'a}&=&
\int_{-\infty}^{\infty}\d x L_{a'}(x)\partial_x L_a(x)\\
&=&\left\{\begin{array}{lll}0&&,\,a'=a\\
&&\\ 
\frac{(-1)^{a'-a}}{a'-a} &&,\,a'\neq a\,,
   \end{array} \right. \nonumber
\end{eqnarray} 
and
\begin{equation}\nonumber
 \left(\partial_x^2\right)_{a'a}=
\left\{\begin{array}{lll}-\frac{\pi^2}{3}&&,\,a'=a\\
&&\\ 
-\frac{2(-1)^{a'-a}}{(a'-a)^{2}}&&,\,a'\neq a\,.
   \end{array} \right.
\end{equation} 
The potential energy matrix elements can be calculated analytically for some
potentials, but it turns out that for the smooth potentials, these can be accurately approximated 
using the Gauss quadrature
of Eq.~\eqref{GaussQuadrature} as
\begin{equation}\nonumber
 V_{a'a}\approx V(a)\delta_{a'a}\,.
\end{equation} 
Strictly speaking, this approximation breaks the variational principle. The validity of the Gauss quadrature approximation 
is discussed in the Appendix. 

Generalized to two dimensions and an area
\mbox{$(-\frac{L}{2},\frac{L}{2})\times(-\frac{L}{2},\frac{L}{2})$},
the Lagrange-sinc functions are given by
\begin{equation}\label{SincMesh2D}
 L_{\va}(\rr)=\tfrac{N}{L}\sinc\left[\tfrac{N}{L}(x-x_{a_x})\right]\sinc\left[\tfrac{N}{L}(y-y_{a_y})\right]\,,
\end{equation} 
where the $N\times N$ mesh points are scaled to 
\begin{equation}\nonumber
\rr_{\va}=\tfrac{L}{N}\va \, , \ \ a_x,a_y \in \left\{-\tfrac{N-1}{2}\,,\,-\tfrac{N-1}{2}+1\,,\,\ldots\,,\,\tfrac{N-1}{2}\right\}\,,
\end{equation} 
having grid spacing $h=L/N$ and weights $\lambda_{\va}=h^2$. The
matrix elements of the Hamiltonian (\ref{eq:h1}) between the basis functions in Eq.~\eqref{SincMesh2D}, in effective atomic units,
are
\begin{align}\nonumber
&H_{a_x'a_y'a_xa_y}=
&\left\{\begin{array}{ll}
\frac{\pi^2}{3 h^2} + V(\rr_{\va})&,\,\text{\begin{footnotesize}$a_x'=a_x,a_y'=a_y$\end{footnotesize}}\\&\\
\frac{(-1)^{a_y'-a_y}}{h^2 (a_y'-a_y)^2}&,\,\text{\begin{footnotesize}$a_x'=a_x,a_y'\neq a_y$\end{footnotesize}} \\&\\
\frac{(-1)^{a_x'-a_x}}{h^2 (a_x'-a_x)^2}&,\,\text{\begin{footnotesize}$a_x'\neq a_x,a_y'= a_y$\end{footnotesize}} \\&\\
0&,\,\text{\begin{footnotesize}$a_x'\neq a_x,a_y'\neq a_y\, ,$\end{footnotesize}} 
           \end{array}\right. \nonumber
\end{align}
where $V(\rr)$ is the external potential. Diagonalization of this Hamiltonian matrix
gives the one-particle eigenstates and -energies. The accuracy of the results obtained can be tested
by varying the number of mesh functions $N$ and the side length of the simulation square $L$.

\subsection{Many particles}

After the one-particle eigenstates are obtained, these can be used as the single-particle 
basis for solving the eigenstates of the interacting
many-body system by exact diagonalization. The $N$-particle Hamiltonian can be
written in the second quantization formalism as
\begin{eqnarray}
 \hat{H}&=&\sum_{j}\varepsilon_{j}\cre_j\ann_j+\frac12\sum_{i,j,k,l}V_{ijkl}\cre_i\cre_j\ann_l\ann_k\ \nonumber \\
 &+&\sum_{i,j}V_{i,j}\cre_i\ann_j,\label{2nd Q Hamiltonian}
\end{eqnarray} 
where $\varepsilon_{j}$ 
are the energy eigenvalues of the single-particle Hamiltonian,
\begin{equation}\label{interaction matrix elements}
V_{ijkl}=\braket{\psi_i\psi_j}{\hat V(\rr,\rr')}{\psi_k\psi_l}
\end{equation}
are the matrix elements of the Coulomb two-body interaction
$\hat{V}(\rr,\rr')=1/|\rr-\rr'|$ in the single-particle basis, and
$V_{i,j}=\braket{\psi_i}{\hat{V}(\v{r})}{\psi_j}$, where $\hat{V}(\v{r})$ contains
the Zeeman interaction and additional external potentials that are not included in Eq. (\ref{eq:h1}). 

The interaction matrix elements of Eq.~\eqref{interaction matrix elements} can
be calculated as follows. Let $\psi_i$ be the single-particle eigenfunctions
expanded in the sinc basis of Eq.~\eqref{SincMesh2D},
\begin{equation}\nonumber
 \psi_i(\rr)=\sum_{\va}\alpha_{\va}^iL_{\va}(\rr)\ .
\end{equation} 
The interaction matrix elements are then
\begin{equation}
\begin{aligned}
 V_{ijkl}&=\int_{\mathbb{R}^2}\!\!\d\rr_1\int_{\mathbb{R}^2}\!\!\d\rr_2 \Psi_i^*(\rr_1)\Psi_j^*(\rr_2)\frac{1}{r_{12}}\Psi_k(\rr_1)\Psi_l(\rr_2)\nonumber\\
&=\sum_{\va,\vb,\vc,\vd}\alpha_{\va}^{i*}\alpha_{\vb}^{j*}\alpha_{\vc}^k\alpha_{\vd}^l 
\int_{\mathbb{R}^2}\!\!\d\rr_1\int_{\mathbb{R}^2}\!\!\d\rr_2\nonumber\\
&\times
L_{\va}(\rr_1)L_{\vb}(\rr_2)\frac{1}{r_{12}}L_{\vc}(\rr_1)L_{\vd}(\rr_2) \label{bc} \nonumber \\
&=\sum_{\va,\vb,\vc,\vd}\alpha_{\va}^{i*}\alpha_{\vb}^{j*}\alpha_{\vc}^k\alpha_{\vd}^l v_{\va\vb\vc\vd} \ ,
\end{aligned} 
\end{equation}
where the expansion coefficients $\alpha$ multiply the interaction matrix
elements $v_{\va\vb\vc\vd}$ between the sinc basis functions. To calculate these, we start with the two-dimensional
Fourier transform of ${1}/{r_{12}}$, namely
\begin{equation}
\begin{aligned}\label{eq:FourierTransf}
 \mathcal{F}\left[\frac{1}{r_{12}}\right](\kk)&=\int_{\mathbb{R}^2}\d\rr_{12}\frac{e^{-i\kk\cdot\rr_{12}}}{r_{12}}\\ \nonumber
&=\int_0^{\infty}\d r_{12}\int_0^{2\pi}\d\theta e^{ikr_{12}\cos(\theta+\pi)}\,,
\end{aligned} 
\end{equation}
where $\theta$ is the angle between $\rr_{12}$ and $\kk$. Using the Jacobi-Anger identity of Bessel functions, 
\begin{equation}\nonumber
  e^{iz\cos\phi}=\sum_{n=-\infty}^{\infty}i^nJ_n(z)e^{in\phi}\,,
\end{equation}
leads to
\begin{equation}
\begin{aligned}
 \mathcal{F}\left[\frac{1}{r_{12}}\right](\kk)&=\int_0^{\infty}\!\!\!\!\d r_{12}\int_0^{2\pi}\!\!\!\!\d\theta\sum_{n=-\infty}^{\infty}(-i)^nJ_n(kr_{12})e^{in\theta}\nonumber\\
&=2\pi\int_0^{\infty}\d r_{12}J_0(kr_{12})=\frac{2\pi}{k}\,.
\end{aligned}
\end{equation}
The potential ${1}/{r_{12}}$ can now be written as the inverse Fourier
transform of $\mathcal{F}\left[\frac{1}{r_{12}}\right]$ as:
\begin{eqnarray}
 \frac{1}{r_{12}}&=&\left(\mathcal{F}^{-1}\circ\mathcal{F}\right)\left[\frac{1}{r_{12}}\right]=\frac{1}{(2\pi)^2}\int_{\mathbb{R}^2}\dd\kk \frac{2\pi}{k}e^{i\kk\cdot\rr_{12}}\nonumber\\
&=&\frac{1}{2\pi}\int_{\mathbb{R}^2}\d k_x\d k_y\,\frac{1}{k}\,e^{ik_x(x_2-x_1)}e^{ik_y(y_2-y_1)}\,. \label{InvFourierTransf}
\end{eqnarray} 
With the identity of Eq.~\eqref{InvFourierTransf}, the integrations over different coordinates factorize in the interaction matrix element:
\begin{eqnarray}
&v_{\va\vb\vc\vd}&=\frac{N}{2\pi L}\int_{\mathbb{R}^2}\d\kk\,\frac{1}{k}\nonumber\\
&\times&\int_{-\infty}^{\infty}\d x_1 \sinc(x_1-a_x)\sinc(x_1-c_x)e^{ik_xx_1}\nonumber\\
&\times&\int_{-\infty}^{\infty}\d y_1 \sinc(y_1-a_y)\sinc(y_1-c_y)e^{ik_yy_1}\nonumber\\
&\times&\int_{-\infty}^{\infty}\d x_2 \sinc(x_2-b_x)\sinc(x_2-d_x)e^{-ik_xx_2}\nonumber\\
&\times&\int_{-\infty}^{\infty}\d y_2 \sinc(y_2-b_y)\sinc(y_2-d_y)e^{-ik_yy_2}.\label{vabcd2}
\end{eqnarray} 
The sinc functions can be replaced by their integral representation
\begin{equation}\nonumber
\sinc(x)=\frac{1}{2\pi}\int_{-\pi}^{\pi}\!\!\!\d t\;e^{ixt}\,,
\end{equation}
and the integrals over $x$ and $y$ coordinates are of the form
\begin{align}
& I_{ab}(k)=\int_{-\infty}^{\infty}\d x \sinc(x-a)\sinc(x-b)e^{ikx}\nonumber\\
&=\left\{\begin{array}{ll}
 \frac{i\sign(k)}{2\pi(a-b)}(-1)^{a-b}\left(e^{ika}-e^{ikb}\right)& ,\;|k|\leq 2\pi,\;a\neq b\\&\\
 \frac{1}{2\pi}e^{ika}(2\pi-|k|)& ,\;|k|\leq 2\pi,\;a=b\\&\\
 0 & ,\;|k|>2\pi\,. 
\end{array}\right.\nonumber
\end{align}
By substituting this result into Eq.~\eqref{vabcd2}, the
original four-dimensional integral over two planes
reduces into a two-dimensional integral over a finite
square in k-space,
\begin{eqnarray}
 v_{\va\vb\vc\vd}=2
&&\int_{-2 \pi}^{2 \pi}\d k_x  \int_{-2 \pi}^{2 \pi}  \d k_y \frac{1}{k} \nonumber \\
&&\times
I_{a_xc_x}(k_x) 
I_{a_yc_y}(k_y) 
I_{b_xd_x}(k_x) 
I_{b_yd_y}(k_y) \nonumber\\
&=&\int_0^{2\pi} \d \theta\int_0^{K(\theta)} \d k \nonumber \\
&&\times I_{a_xc_x}(-k \cos(\theta)) 
I_{a_yc_y}(-k \sin(\theta)) \nonumber \\
&&\times I_{b_xd_x}(k \cos(\theta)) 
I_{b_yd_y}(k \sin(\theta)) \ ,
\label{nelio}
\end{eqnarray}
where $K(\theta)=2\pi/\max(|\cos(\theta)|,|\sin(\theta)|)$ is the radial
integration limit corresponding to the square. The last form can be used in
numerical calculations.

One can see that in Eq.~\eqref{nelio}, one obtains five different
integrals depending on how many of the four functions $I_{ab}$ have
the same indices. In addition, the case with two equal index pairs is
naturally split into two cases, depending on whether the equal indices
belong to the same Cartesian component of $k$.
 In most cases, some further analytic work can be done to handle the
angular integral. For instance, in the case when all the index pairs
differ, such that $a_x\neq c_x$, $b_x\neq d_x$, $a_y\neq c_y$ and
$b_y\neq d_y$, the integrand can be written as a sum of terms of the
form $ \cos\left\{k[m \cos(\theta)+ n \sin(\theta)]\right\} $, and the
angular part can be integrated analytically, and we are left with a
one-dimensional numerical integral.  In this way, we are able to
calculate the interaction matrix elements between the sinc basis
functions, and then for any external confinement
potential, Eq.~\eqref{bc} can be used to construct $V_{ijkl}$.

It turns out that the calculation of $V_{ijkl}$ from Eq.~\eqref{nelio} is
computationally very time-consuming, because one has to loop over four
indices on both the right- and left-hand sides of Eq.~\eqref{nelio}.
Luckily, this basis change can be trivially parallelized and a very
efficient scheme can be obtained using graphics processing units (GPUs).

We performed the calculation of the interaction elements in Eq.~\eqref{nelio} with an Nvidia Tesla C2070 graphics processing unit, which was programmed with CUDA\cite{CUDA}, a parallel programming model for Nvidia GPUs. On the GPU, the computation is parallelized across tens of thousands of lightweight computational threads, which are organized in independent blocks. In our parallelization scheme, each block computes one element of $V_{ijkl}$. Inside the block, the sum over the index $\mathbf{d}$ is parallelized across the threads with each thread corresponding to a value of $\mathbf{d}$. The threads then loop over the indices $\mathbf{a},\mathbf{b}$ and $\mathbf{c}$, and in the end the results of all threads in the block are summed with a parallel prefix sum algorithm to obtain the final result.
 
In Eq.~\eqref{nelio}, $v_{\mathbf{abcd}}$ does not depend on the state indices $i,j,k,l$, and it is beneficial to calculate it beforehand and store it in a table in the GPU memory. In double precision floating point arithmetic, the size of the table for a $N\times N$ mesh is $8N^8$ bytes. For a $12\times 12$ mesh, the size is approximately $3.3$ gigabytes, which fits into the $6$ gigabyte global memory of the state of the art Tesla cards, such as the C2070. We also utilize the fast on-chip shared memory by caching the expansion coefficients $\alpha$ before the calculation. The GPU speeds up the matrix element calculation by a factor of around 13 when double precision arithmetic is used.

Unfortunately, we have found the $12\times 12$ mesh insufficient for double quantum dot calculations if a realistic distance between the minima is used. Therefore, the calculation has to be divided so that the whole $v_{\mathbf{abcd}}$ matrix is not calculated at once. We lowered the memory requirement by calculating first $V_{ijkl}^{a_x}$ by fixing the $a_x$ index in Eq.~\eqref{nelio}. As a consequence, it is sufficient to calculate the $v_{\mathbf{abcd}}$ also using a fixed $a_x$ index, and the memory requirement is dropped to $8N^7$ bytes, which allows calculation with a $17\times 17$ mesh. The $V_{ijkl}$ elements are obtained by summing the $V_{ijkl}^{a_x}$ elements over $a_x$. The sum is updated after the calculation of each $V_{ijkl}^{a_x}$ matrix to save memory. This modification of the algorithm adds some serial work, which slows down the computation, but a compromise between memory requirement and speed must be made.

The sum in Eq.~\eqref{nelio} could be further divided to allow larger mesh sizes by fixing more indices, but the computation time becomes fast a limiting factor. The calculation of interaction matrix elements for the 24 lowest single-particle states using a mesh size of $17\times 17$ takes almost two days. The exact diagonalization part is much faster than this.

\section{Results}

\subsection{Convergence of the single particle states}

In this section, we compute the single-particle eigenstates of systems consisting of
1, 2, and 4 minima using the Lagrange mesh method. The accuracy of the method and optimal parameters are discussed.
In the following sections, we then use the obtained single particle states in the actual many-body computations. 

\begin{figure}[!ht]
\vspace{0.3cm}
\includegraphics[width=0.90\columnwidth]{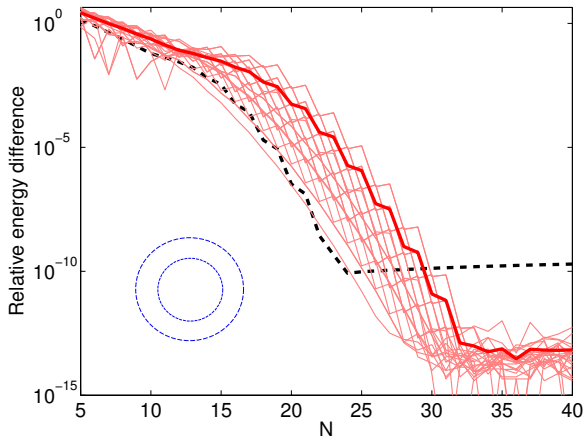}
\caption{(Color online) The convergence of the 24 first single particle energies in the case of one parabolic dot as a function of $N$ (the grid being $N\times N$).
The relative differences of the single-particle energies (computed with the Lagrange mesh method) with the analytical Fock-Darwin energies are shown. 
The thin red curves show the first 24 states case with $L=280$ nm (the thick red curve shows the average relative difference). The thick dashed curve shows the average relative differences in the smaller grid case with $L=200$ nm.}
\label{fig:one}
\end{figure}

First, we studied the convergence of the method in the analytically solvable case of just one parabolic well. The confinement strength
was $\hbar\omega_0=4$ meV. We computed the 24 first single
particle energies with different mesh parameters $N$ and $L$ and compared the results with the analytical Fock-Darwin eigenenergies. The relative difference of the energies
can be seen as a function of the grid size $N$ in Fig. \ref{fig:one}.

Fig. \ref{fig:one} shows that given large enough $N$, the relative difference of the energies converges to the order of the numerical double precision
accuracy in the $L=280$ nm case. The effect of the size of the simulation area can also be seen in the figure. The smaller area case (the black dashed curve, $L=200$ nm) 
shows faster convergence with respect to $N$. However, the finite simulation area results in some error as well, and thus the convergence in the $L=200$ nm case
stops before it reaches the double precision.

The main topic of this paper is the simulation of singlet-triplet qubits. We will first study one-qubit dynamics and then use our model to simulate a system of two singlet-triplet qubits.
Next we discuss the convergence of the method in these systems.

In the potential in Eq. (\ref{eq:qdpot}), the derivative of the potential is not continuous; the $\min$-function causes an edge at the interface of two branches.
This sharp edge can be problematic in the Lagrange mesh method due to the finite number of mesh points. To alleviate this, rounding of the edges was used
in the case of multiple dots. The rounding was found to speed up the convergence of the single particle states.

\begin{figure}[!ht]
\vspace{0.3cm}
\includegraphics[width=\columnwidth]{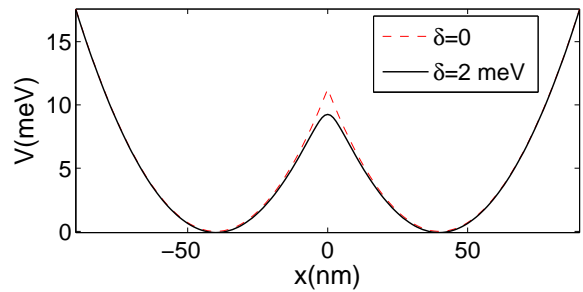}
\caption{(Color online) The effect of the rounding on the DQD-potential. The potential is shown in the $x$-axis. The minima are located at $x=\pm40$ nm. The confinement strength is $\hbar\omega_0=4$ meV.
Both non-rounded ($\delta=0$, red dashed curve) and rounded ($\delta=2$ meV, black solid curve) potentials are shown.
}
\label{fig:rounding}
\end{figure}

The rounding is achieved by defining a matrix $\mathbf{R}$ at each grid point. $\mathbf{R}$ has the different dot potentials in its diagonal. For example, in the case of four dots at locations
$\v{r}_1...\v{r}_4$ the diagonal entries are $R_{11}=V_1=\frac{1}{2}m^*\omega_0^2|\v{r}-\v{r}_1|^2$ and so on. The non-diagonal entries are constant $\delta$ and define the strength of the rounding. The potential at
the particular grid point is given as the smallest eigenvalue of $\mathbf{R}$. The effect of the rounding can be seen in Fig. \ref{fig:rounding}.

The current maximum grid size in the computation of the $V_{ijkl}$-elements is $N=17$ due to the GPU memory limitations (larger grids can in principle be computed, but with the expense
of considerably longer computations times).
As the accuracy of the method depends non-trivially on both the simulation area $L$ and the grid size $N$, the value of $L$ was optimized.

\begin{figure}[!ht]
\vspace{0.3cm}
\includegraphics[width=0.90\columnwidth]{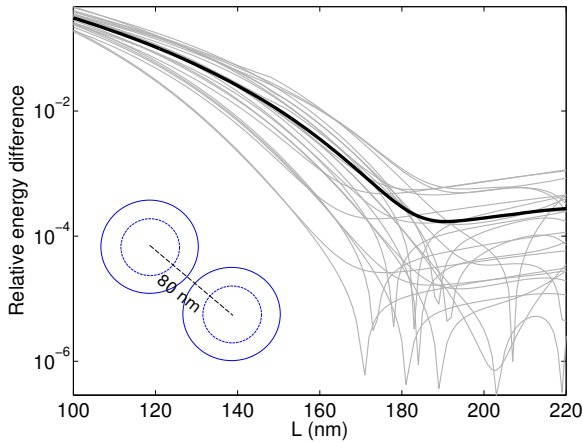}
\caption{The convergence of the 24 first Lagrange mesh single particle energies as a function of the simulation area length $L$ in the two dot case.
The two-dot potential is illustrated in the inset. The rounding is set to $\delta=2$ meV. The grid size is $17\times17$ ($N=17$).
The relative differences of the Lagrange mesh energies with the energies of a larger
system ($L=300$ nm and $N=68$, also computed with the Lagrange mesh method) are shown. The thick curve shows the average relative difference for the 24 states.
}
\label{fig:two}
\end{figure}

\begin{figure}[!ht]
\vspace{0.3cm}
\includegraphics[width=0.90\columnwidth]{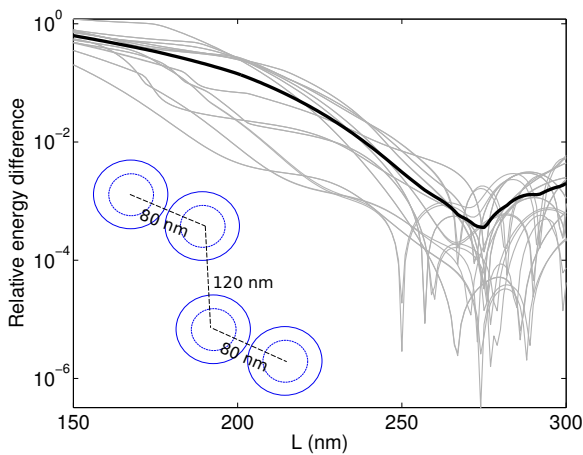}
\caption{The same as in Fig. \ref{fig:two} but for the four dot case, with the reference system being $L=320$ nm and $N=68$.
}
\label{fig:four}
\end{figure}

We compared the obtained eigenenergies with those of a large system ($N=68$ and $L=300$ nm or $L=320$ nm) and chose the value for $L$ that gave the smallest error
with respect to the more accurate large system.
The relative difference of the energies as a function of $L$ in the two dot case is shown in Fig. \ref{fig:two} and in Fig. \ref{fig:four} in the four-dot case.
The potentials for the two- and four-dot systems are illustrated in the insets of Figs. \ref{fig:two} and \ref{fig:four}.
The two-dot potential consists of parabolic dots with the distance $80$ nm between their minima. In the four-dot system, dots $1$ and $2$ are $80$ nm apart, dots $2$ and $3$ are $120$ nm apart, and $3$ and $4$ are $80$ nm apart.

Figs. \ref{fig:two} and \ref{fig:four} show that with $N=17$ the optimal value of $L$ is between $180$ nm and $200$ nm in the two-dot case and between $260$ nm and $290$ nm in the four-minima case. Up to this point, the convergence of the energies
is monotonous. With too small $L$, the wave function 'leaks' out of the simulation area, and with too high $L$ the grid spacing becomes too large.
The singularity like dips in the relative difference curves probably result from the fact that the errors due to finite $L$ and $N$ have different signs.
At the dip, these errors nearly cancel each other out.

\FloatBarrier

\subsection{Singlet-triplet qubit}

In this section, we use the Lagrange mesh and ED methods to simulate the time evolution of the state of a singlet-triplet DQD qubit.
We demonstrate that, by applying local electric and magnetic fields in our model, we can achieve full quantum control
over the state of the qubit and reproduce realistic dynamics of the system in our simulation.

We used a potential that consists of two parabolic dots,
\begin{equation}
V(\v{r})=\frac{1}{2}m^*\omega_0\min\{|\v{r}-\v{r}_1|^2,|\v{r}-\v{r}_2|^2\},
\end{equation}
to model a singlet-triplet qubit. The confinement strength was $\hbar\omega_0=4$ meV and the distance between the dots was $a=|\v{r}_1-\v{r}_2|=80$ nm.
Our DQD potential is illustrated in the inset of Fig. \ref{fig:two}.

The logical basis of a singlet-triplet qubit consists of the two lowest eigenstates,
the singlet state, $|S\rangle=\frac{1}{\sqrt{2}}(|\uparrow\downarrow\rangle-|\downarrow\uparrow\rangle)$, 
and the $S_z=0$ triplet state, $|T_0\rangle=\frac{1}{\sqrt{2}}(|\uparrow\downarrow\rangle+|\downarrow\uparrow\rangle)$
(the arrows denote direction of the electron spins).
An arbitrary state $|\psi\rangle$ of the qubit can be written as
\begin{equation}\label{eq:qubit}
|\psi\rangle=\cos\left(\frac{\theta}{2}\right)|S\rangle+\sin\left(\frac{\theta}{2}\right)e^{i\phi}|T_0\rangle.
\end{equation}
Here, $\theta\in\left[0,\pi\right]$ and $\phi\in\left[0,2\pi\right)$. The state of the qubit can thus be visualized using the surface of a Bloch sphere
with $|S\rangle$ and $|T_0\rangle$ at the north and south poles, and $\theta$ and $\phi$ denoting the angles with respect to $z$ and $x$-axes.

It should be noted that the DQD is not a true two-level system. There are higher excited states as well, and 
Eq. (\ref{eq:qubit}) is just an approximation. However, in our simulations, the weighs of the higher states were found to be negligible with practical parameter values,
and the system can be considered as two level in this sense.

Universal quantum control of the qubit requires rotations around at least two different axes in the aforementioned Bloch sphere.
In DQD singlet-triplet qubits, rotations around the $z$-axis are controlled by the exchange interaction (the singlet-triplet energy difference) $J=E_{T_0}-E_{S}$ and rotations around the $x$-axis
can be generated by a magnetic field gradient $\Delta B_z$ between the dots. The axis of the rotation in the Bloch sphere is then 
\begin{equation}\label{eq:axis}
\v{n}=J\hat{\v{e}}_z+\mu_Bg\Delta B_z\hat{\v{e}}_x,
\end{equation}
and the frequency is\cite{foletti}
\begin{equation}\label{eq:freq}
f=\sqrt{J^2+(\mu_Bg\Delta B_z)^2}/h.
\end{equation}
Here, $g=-0.44$ is the GaAs gyro magnetic ratio, $\mu_B$ the Bohr magneton and $h$ the Planck's constant.

The single particle eigenstates are computed using the Lagrange mesh method and they are then used
in the two particle ED-calculations. In our model, the $z$-rotations are created by detuning (a potential energy difference $\epsilon$ between the minima of the dots)
the two parabolic dots, which lifts the degeneracy
of the $|S\rangle$ and $|T_0\rangle$ states and results in exchange interaction. The $x$-rotations are created using a local magnetic field gradient that
is taken into account by the Zeeman-term.

The detuning potential and the local magnetic field are modeled as step functions that are zero far away
from the dot minima and have different signs in the two dots. We calculate the the matrix elements 
$V_{i,j,\sigma_1,\sigma_2}=\langle\psi_{i,\sigma_1}|V|\psi_{j,\sigma_2}\rangle$, where $V$ is either the detuning potential or the Zeeman-interaction,
in the eigenbasis $\{\psi_{i,\sigma}\}$ obtained using the Lagrange mesh method ($\sigma$ denotes the spin quantum number). 
The detuning and the Zeeman-term are then taken into account in the two-body ED through the one-body operator $\hat{V}=\sum_{i,j,\sigma_1,\sigma_2}V_{i,j,\sigma_1,\sigma_2}a^{\dagger}_{i,\sigma_1}a_{j,\sigma_2}$.

The evolution of the initial state $|\psi(0)\rangle$ of the qubit is computed by propagation, using
\begin{equation}\label{eq:prop}
|\psi(t+\Delta t)\rangle=\exp\left(-\frac{i}{\hbar}\hat{H}(t)\Delta t\right)|\psi(t)\rangle,
\end{equation}
where $\hat{H}(t)$ is the (time-dependent) two-body Hamiltonian. The matrix exponent is computed using Lanczos method.
To study the evolution of the qubit's state in the Bloch sphere,
the angles $\theta$ and $\phi$ in (\ref{eq:qubit}) are extracted from $|\psi(t)\rangle$ by using the properties of the two-body
spin operator $\hat{S}^2$, i.e. $\hat{S}^2|S\rangle=0$ and $\hat{S}^2|T_0\rangle=2\hbar^2|T_0\rangle$.

The first 24 single-particle states were computed using the Lagrange mesh method. The mesh parameters were $N=17$ and $L=210$ nm. The rounding was set to $\delta=2$ meV.
The $V_{ijkl}$- and $V_{ij}$-elements (corresponding to both the detuning and the Zeeman term) were computed for the 24 single-particle states.

We first demonstrate the control of the qubit in a simple case. In this simulation, the system is initially in the singlet state.
The dots are detuned so that the difference between their energy minima is $\epsilon=V_2-V_1=4.3$ meV.
The detuning lifts the degeneracy of the singlet
and triplet states, resulting in an exchange energy of $J\approx3.748$ $\mu$eV.
A magnetic field difference of $\Delta B_z=0.4$ is then put between the dots and the system is let to evolve for $1$ ns. The singlet and triplet probabilities
were computed by projecting the state of the qubit onto the $S^2$ operator.

\begin{figure}[!ht]
\vspace{0.3cm}
\includegraphics[width=\columnwidth]{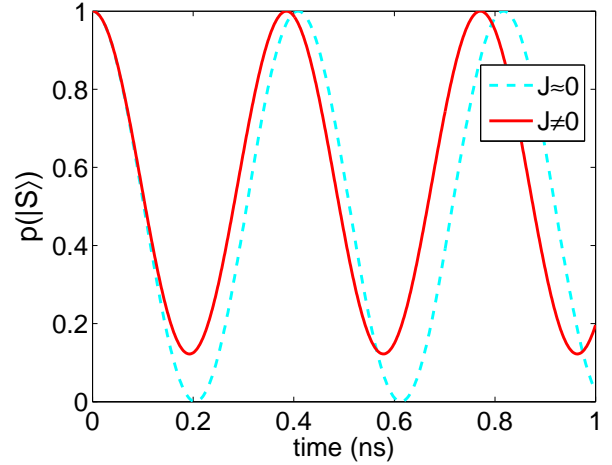}
\caption{(Color online) The time evolution of the singlet probability $p(|S\rangle)$ (red curve). The detuning is $\epsilon=4.3$ meV, and the magnetic field is $\Delta B_z=0.4$ T.
This part is omitted from the figure, as $p(|S\rangle)$ is constant during it. The simulation time is $1$ ns and the time step length $\Delta t=1$ ps.
The dashed cyan curve shows the non-detuned case with $J\approx0$.}
\label{fig:sprob}
\end{figure}

\begin{figure}[!ht]
\vspace{0.3cm}
\includegraphics[width=\columnwidth]{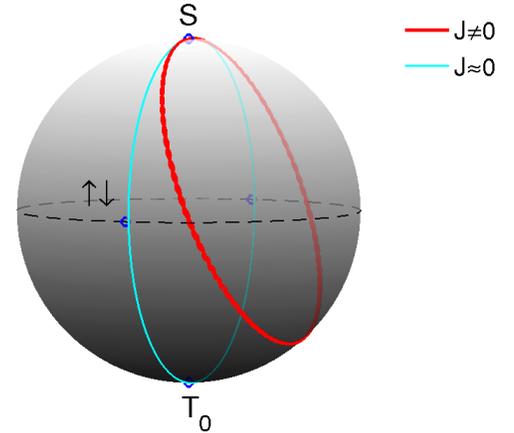}
\caption{(Color online) The evolution of the state of the DQD qubit on the Bloch sphere. The qubit is initially in the singlet state (at the north pole). A detuning of $\epsilon=4.3$ meV and a magnetic field gradient of $\Delta B_z=0.4$ are turned on and
the state is let to evolve. The simulation time is $1$ ns, and $\Delta t$=1 ps. The thick red curve denotes the trajectory of the qubit's state on the sphere. The thin cyan curve shows the non-detuned case, with $J\approx0$. The dashed black curve (the equator) shows the $xy$-plane, where
$p(| S \rangle)=1/2$.}
\label{fig:bloch}
\end{figure}

The computed time evolution of the singlet
probabilities $p(|S\rangle)$ can be seen in Fig. \ref{fig:sprob}. Fig. \ref{fig:bloch} shows the evolution of the state of the qubit on the
Bloch sphere. 

In Fig \ref{fig:sprob}, the detuned singlet probability oscillates between its maximum $1$ and minimum $0.12$. The singlet probability never goes to zero
due to the $z$-rotation driven by the exchange energy $J\approx3.748$ $\mu$eV.  
The frequency of the oscillation is
$f\approx2.618$ GHz, which is very close to the value given by Eq. (\ref{eq:freq}), $f\approx2.625$ GHz. In the non-detuned case, the probability oscillates between $1$ and $0$, as expected.
In this case too, the computed frequency coincides very well with Eq. (\ref{eq:freq}).

Fig. \ref{fig:bloch} shows that in the detuned case, the plane of the rotation is tilted from the $|S\rangle$-$|T_0\rangle$ plane, as expected by Eq. (\ref{eq:axis}).
The state never reaches $|T_0\rangle$ (the south pole) during the simulation. The non-detuned case oscillates between $|S\rangle$ and $|T_0\rangle$, passing through
the spin localized states $|\uparrow\downarrow\rangle=1/\sqrt{2}(|S\rangle+|T_0\rangle)$ and $|\downarrow\uparrow\rangle=1/\sqrt{2}(|S\rangle-|T_0\rangle)$.

\begin{figure}[!ht]
\vspace{0.3cm}
\includegraphics[width=\columnwidth]{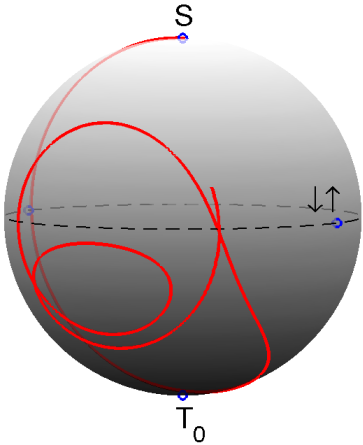}
\caption{(Color online)  evolution of the state of the DQD qubit on the Bloch sphere. The qubit is initially in the singlet state (at the north pole). An oscillating detuning,
 $\epsilon(t)=\epsilon_0\sin(2\pi ft)$, where $\epsilon_0=4.6$ meV and $f=1$ GHz, and magnetic field $\Delta B_z=0.4$ T, are then applied. The simulation time is $1$ ns, and $\Delta t=1$ ps.}
\label{fig:bloch_hullu}
\end{figure}

We also tried more complicated pulse sequences and tracked the evolution of the state in the Bloch sphere. One such is demonstrated in Fig. \ref{fig:bloch_hullu}.
Here, the detuning strength was oscillating, $\epsilon(t)=\epsilon_0\sin(2\pi ft)$, where $\epsilon_0=4.6$ meV and $f=1$ GHz. A non-trivially time dependent $J$ causes
the axis of the states's rotation to change as a function of time, which leads to quite complicated paths on the Bloch sphere.

In conclusion, our model, based on Lagrange mesh ED, can be used to simulate GaAs singlet-triplet DQD qubits. We can simulate the realistic full quantum
control of the qubit starting from the first principles. Next, we proceed to use the model in studying two-qubit dynamics.
\FloatBarrier

\subsection{Two singlet-triplet qubits}

The Lagrange mesh allows the study of interplay of singlet-triplet qubits. For example, the entanglement of two singlet-triplet qubits by the long distance dipole-dipole interaction can be simulated using this method. 
In this section, we first compute and study the lowest eigenstates of the two-DQD system, using different detunings for the two qubits. The main topic of this section is the entanglement of
singlet-triplet qubits. We show that our model can be used to simulate the entangling procedure demonstrated recently by Shulman et al\cite{shulman}. 

\begin{figure}[!h]
\vspace{0.3cm}
\includegraphics[width=0.90\columnwidth]{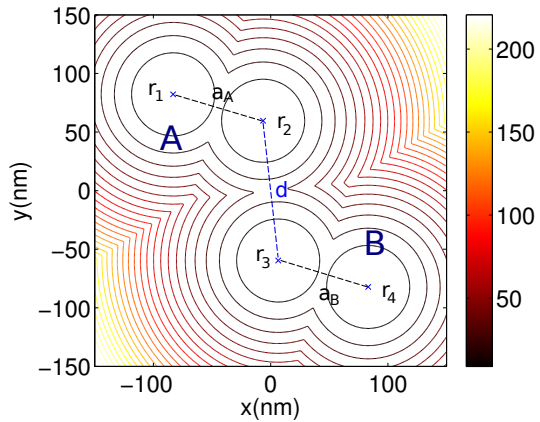}
\caption{Contour plot of the two-DQD potential. The system consists of four QDs, divided to
two qubits A and B. The confinement strength is $\hbar\omega_0=4$ meV. The qubit-qubit distance is $d=120$ nm, and the distance of the dots in the qubits
are $a_A=a_B=80$ nm. The contours are shown in meV.
}
\label{fig:fourdots}
\end{figure}

We model the two-DQD system with an external confinement potential that is the minimum of four quadratic wells,
\begin{equation}
V(\v{r})=\frac{1}{2}m^*\omega_0^2\min_{j \leq 4}\{|\v{r}-\v{r}_j|^2\}.
\end{equation}
Our simulation system can be divided to qubits A and B. A consists of the wells at $\v{r}_1$ and $\v{r}_2$, with the dot distance
$a_A=|\v{r}_1-\v{r}_2|=80$ nm. Similarly, the inter dot distance of the qubit B is $a_B=|\v{r}_3-\v{r}_4|=80$ nm.  The inter qubit distance is given by $d=|\v{r}_2-\v{r}_3|=120$ nm. The confinement strength is $\hbar\omega_0=4$ meV.
The potential is illustrated in Fig. \ref{fig:fourdots}.
The inter-qubit distance and the confinement strangth are large enough that there is no tunneling between A and B, so the qubits interact only through
the Coulomb repulsion of their electrons. Also, the qubits interact mainly via the electrons in the dots 2 and 3, as the inter-dot distances are quite large.

The qubits A and B can become entangled due to the fact that under the exchange interaction, the charge densities of the $|S\rangle$ and $|T_0\rangle$ states differ. When the detuning lowers the potential energy in one of the dots of the qubit, the singlet state charge density becomes more located in this dot. However, if the detuning is not too high the triplet density is unaffected due to the repulsive exchange force in the spatially anti-symmetric triplet state.

The singlet and triplet states have differing charge densities, and 
hence the Coulomb repulsion between the qubits depends on the states of the qubits. This conditioning creates an entangled state when the qubits are evolved under exchange.

A bipartite state $|\psi\rangle_{AB}\in \mathcal{H}_A\otimes \mathcal{H}_B$ ($\mathcal{H}_A$ and $\mathcal{H}_B$ are
the Hilbert spaces of the subsystems $A$ and $B$) is an entangled state if it cannot be written as a tensor product $|\psi\rangle_{AB}=|\psi\rangle_A\otimes|\psi\rangle_B$. 
In general, if the vector $|\psi\rangle_{AB}$ is written in any orthonormal
product basis $\{|e_i\rangle_A\otimes|e_j\rangle_B\}_{ij}$,
\begin{equation}
|\psi\rangle_{AB}=\sum_{i,j}M_{ij}|e_i\rangle_A\otimes|e_j\rangle_B,
\end{equation}
it is an entangled state if and only if the matrix of coefficients, $\mathbf{M}=\{M_{ij}\}$, is not singular.

The degree of entanglement can be determined by some entanglement measure. One such measure is the concurrence. In case of pure states, and two-level systems (qubits) $A$ and $B$, concurrence $C$ is given as $C=2\sqrt{\lambda_1\lambda_2}$, where
$\lambda_1$ and $\lambda_2$ are the eigenvalues of matrix $\mathbf{M}^{\dagger}\mathbf{M}$ . It is easy to see that this simplifies to
the formula
\begin{equation} \label{eq:conc}
C=2|\det(\mathbf{M})|.
\end{equation}
Concurrence can also be generalized to mixed states\cite{ent}.

Concurrence assumes values between $0$ and $1$. A non-zero $C$ is a property of an entangled state, and the higher the value of $C$, the higher the degree of entanglement. The maximally entangled Bell states have $C=1$.

In our two singlet-triplet qubit system, the Hilbert spaces are given as the two lowest eigenstates of a DQD-system, $\mathcal{H}_A=\mathcal{H}_B=\{|S\rangle,|T_0\rangle\}$. The $\mathbf{M}$ matrix is thus obtained by projecting
the four-electron wave function onto the computational basis $\{|SS\rangle,|ST_0\rangle,|T_0S\rangle,|T_0T_0\rangle\}$;  $M_{11}=\langle SS|\psi\rangle$, 
$M_{12}=\langle ST_0|\psi\rangle$, $M_{21}=\langle T_0S|\psi\rangle$ and $M_{22}=\langle T_0T_0|\psi\rangle$.

It should be noted that while the states $|SS\rangle$, $|ST_0\rangle$, and $|T_0S\rangle$ are eigenstates of the four-particle $S^2$ operator,
$|T_0T_0\rangle$ is not. Indeed, it is not given as an eigenstate by the Lanczos iteration. In order to do the projections onto the
computational basis, the state $|T_0T_0\rangle$ was generated using localized magnetic fields.

As $|T_0\rangle=\frac{1}{\sqrt{2}}(|\uparrow\downarrow\rangle-|\downarrow\uparrow\rangle)$, $|T_0T_0\rangle=|T_0\rangle_A\otimes|T_0\rangle_B$ can be written as
\begin{eqnarray}\label{eq:tt}
|T_0T_0\rangle=\frac{1}{2}|\uparrow\downarrow\rangle_A\otimes|\uparrow\downarrow\rangle_B+\frac{1}{2}|\uparrow\downarrow\rangle_A\otimes|\downarrow\uparrow\rangle_B \nonumber \\
+\frac{1}{2}|\downarrow\uparrow\rangle_A\otimes|\uparrow\downarrow\rangle_B+\frac{1}{2}|\downarrow\uparrow\rangle_A\otimes|\downarrow\uparrow\rangle_B. 
\end{eqnarray}
In this decomposition, $|T_0T_0\rangle$ is written using the $S_z=0$ eigenstates. These $S_z=0$ eigenstates can be generated using strong localized magnetic
fields in the four dots of the two-qubit system. For example, $|\uparrow\downarrow\rangle_A\otimes|\uparrow\downarrow\rangle_B$ is obtained as the ground state
of a system where the magnetic field is up in the first dot, down in the second, up in the third and down in the fourth (the Zeeman-term alignes the spins of the electrons along the magnetic fields).

Decompositions similar to Eq. (\ref{eq:tt}) can be written for the other three states as well. Indeed, in the non-detuned case, the singlet states given by such decompositions were found to
be the same eigenstates that Lanczos iteration would find.
However, the aforementioned magnetic field scheme for creating the
$S_z=0$ eigenstates can only be used to create states that have identical density in the two dots of the qubits. Hence, it is not well suited for creating the detuned
singlet states. Fortunately if the detuning is in the practical operation regime of DQD-qubits, the $|T_0T_0\rangle$ density remains symmetric with respect to the two
dots of the qubits.

Thus, the computational basis can be created as follows. The states $|SS\rangle$, $|ST_0\rangle$ and $|T_0S\rangle$ are given as eigenstates by Lanczos and they can be identified
by their spin. The state $|T_0T_0\rangle$ is generated using the decomposition Eq. (\ref{eq:tt}). The wave function can then be projected onto this basis, and the
concurrence can be computed according to Eq. (\ref{eq:conc}).

In the scheme where the qubits A and B are first brought to the $xy$-plane and then let to evolve under exchange (used for example by Shulman et al.\cite{shulman}), we can derive
a simple analytic formula for the time dependence of the concurrence.

In the absence of magnetic fields, the Hamiltonian of the two-qubit system is close to a diagonal one
in the basis $\{|SS\rangle,|ST_0\rangle,|T_0S\rangle,|T_0T_0\rangle\}$ (this was verified numerically). The diagonal entries of the projected Hamiltonian
are the energies $E_{SS}$, $E_{ST_0}$, $E_{T_0S}$ and $E_{T_0T_0}$. As the qubits are let to evolve in the $xy$-plane, the weights in the $\mathbf{M}$ matrix obtain
phase factors proportional to these energies.

Let the system be initially in the state with $M_{ij}=\frac{1}{2}$ $\forall i,j\in\{1,2\}$, i.e. $|\psi\rangle=|\uparrow\downarrow\rangle_A\otimes|\uparrow\downarrow\rangle_B$. The system
is then let to evolve. If we approximate the projected Hamiltonian to be diagonal, the time dependence of the coefficients $M_{ij}$ is given as $M_{ij}(t)=\frac{1}{2}e^{iE_{ij}t/\hbar}$, where $E_{11}=E_{SS}$ and so on.
Inserting these in Eq. (\ref{eq:conc}) yields the formula for the concurrence,
\begin{equation} \label{eq:conc2}
C(t)=\frac{1}{2}\sqrt{2-2\cos\left(\Delta_E t/\hbar\right)},
\end{equation}
where, $\Delta_E=E_{SS}+E_{T_0T_0}-E_{ST_0}-E_{T_0S}$.

In Eq. (\ref{eq:conc2}), the parameter $\Delta_E$ represents the coupling between the qubits. Eq. (\ref{eq:conc2}) shows that the entanglement indeed arises from the differences in the charge densities. If all the computational basis states have
identical charge densities, the Coulomb repulsion between the two qubits, $\gamma$, is the same for all these states.
In this case, the energies of the computational basis states are $E_{SS}=2E_S+\gamma$, $E_{T_0T_0}=2E_{T_0}+\gamma$ and $E_{ST_0}=E_{T_0S}=E_{S}+E_{T_0}+\gamma$.
and $\Delta_E=0$. The concurrence is thus zero, i.e. there is no entanglement between the qubits.  In the detuned case, the states have different densities and different values of the Coulomb repulsion. Hence, $\Delta_E\neq0$, and the concurrence
oscillates according to Eq. (\ref{eq:conc2}).

The single-particle states and the $V_{ijkl}$-elements were again computed using the Lagrange mesh method. The simulation cell area was $L=280$ nm, and the mesh size was $N=17$. The rounding of the edges was set to $\delta=2$ meV. The 24 first one-particle states were used in the four-particle ED computations.
As in the one qubit case, the detuning and local magnetic field matrix elements $V_{i,j,\sigma_1,\sigma_2}=\langle\psi_{i,\sigma_1}|V|\psi_{j,\sigma_2}\rangle$ were computed in order to achieve full quantum control over the qubits.  

First, we study the lowest eigenstates of the four-electron system. Without detuning ($\epsilon_A=V_1-V_2=0$ and $\epsilon_B=V_4-V_3=0$) the six first states (with $S_z=0$) are close to each other in energy. 

\begin{figure*}[!ht]
\vspace{0.3cm}
\includegraphics[width=1.6\columnwidth]{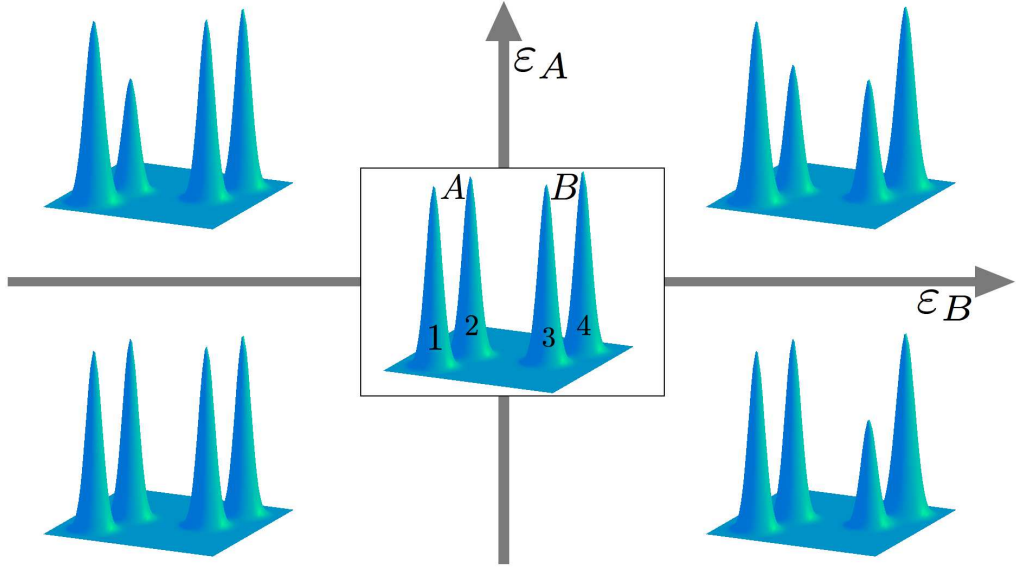}
\caption{The effect of the sign of detuning on the density of the ground state $|SS\rangle$. In the middle: the non-detuned $|SS\rangle$ density. The numbers in the middle plot refer to the dots 1-4 and A and B denote the qubits.
The signs of the detunings, $\epsilon_A=V_2-V_1$ and $\epsilon_B=V_3-V_4$, are shown by the axes $\epsilon_A$ and $\epsilon_B$. In the upper left corner:
$\epsilon_A=4.35$ meV and $\epsilon_B=-4.35$ meV (dots $1$ and $3$ have low potential). In the upper right corner: $\epsilon_A=\epsilon_B=4.35$ meV (dots $1$ and $4$ have low potential).
In the lower left corner: $\epsilon_A=\epsilon_B=-4.35$ meV (dots $2$ and $3$ have low potential). In the lower right corner: $\epsilon_A=-4.35$ meV and $\epsilon_B=4.35$ meV (dots $2$ and $4$ have low potential).}
\label{fig:detuning}
\end{figure*}

The ground state is $|SS\rangle$ ($s=0$), and the next two states are $|ST_0\rangle$ and $|T_0S\rangle$ ($s=1$). The next three states given by Lanczos are what we call the triplet states, superpositions of $s=0$, $s=1$ and $s=2$ eigenstates (in the four-particle case, there
can be several $S^2$ eigenstates with given quantum numbers $s$ and $s_z$)
These three eigenstates of $S^2$ are so degenerate that Lanczos mixes them. The three triplet states share the same energy as $|T_0T_0\rangle$ (which also is not an $S^2$ eigenstate, but another linear combination of the triplet states). The electrons of the first six states are symmetrically located in the four dots, one electron in each.

\begin{figure}[!h]
\vspace{0.3cm}
\subfigure{\includegraphics[width=.49\columnwidth]{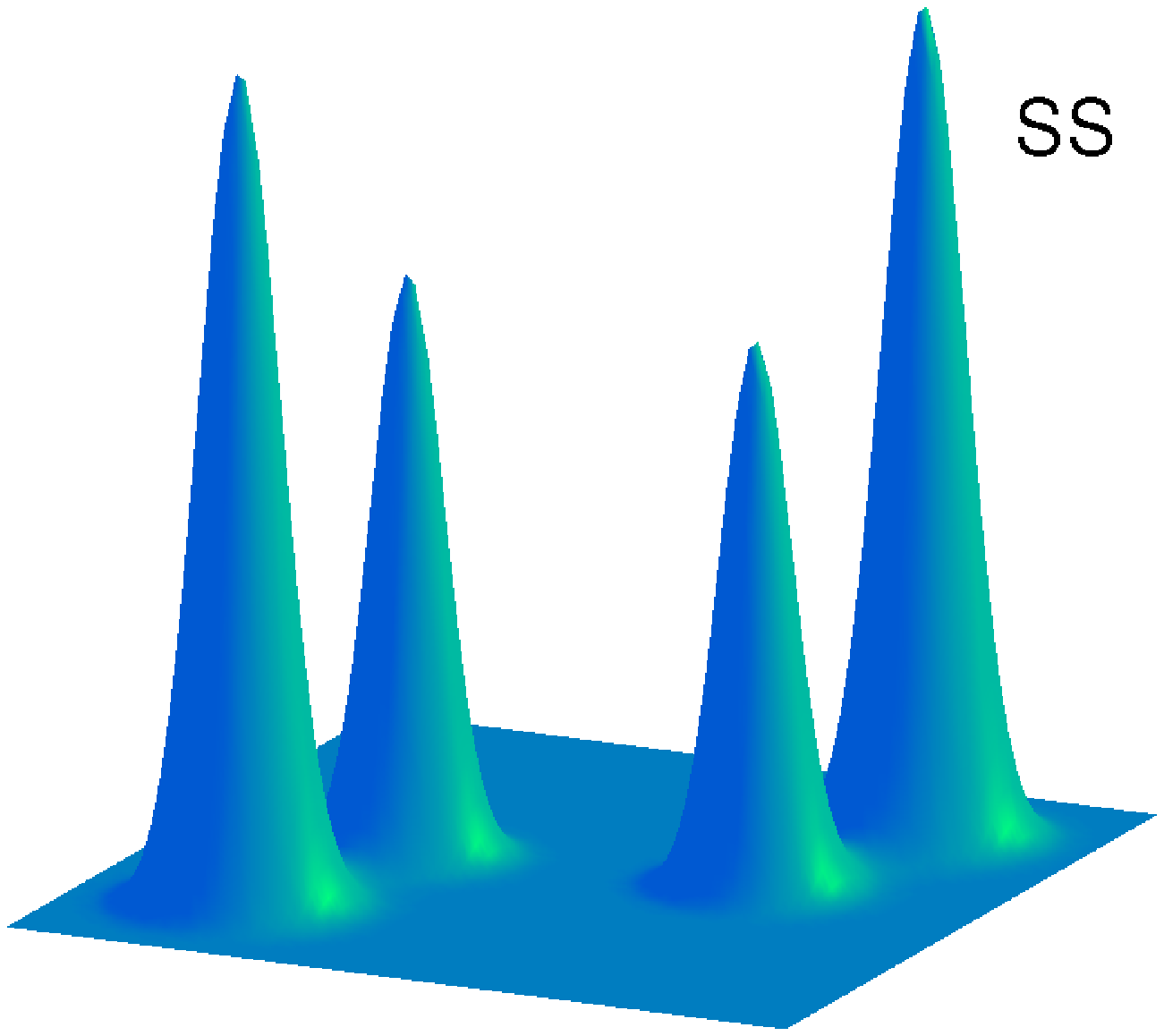}}
\subfigure{\includegraphics[width=.49\columnwidth]{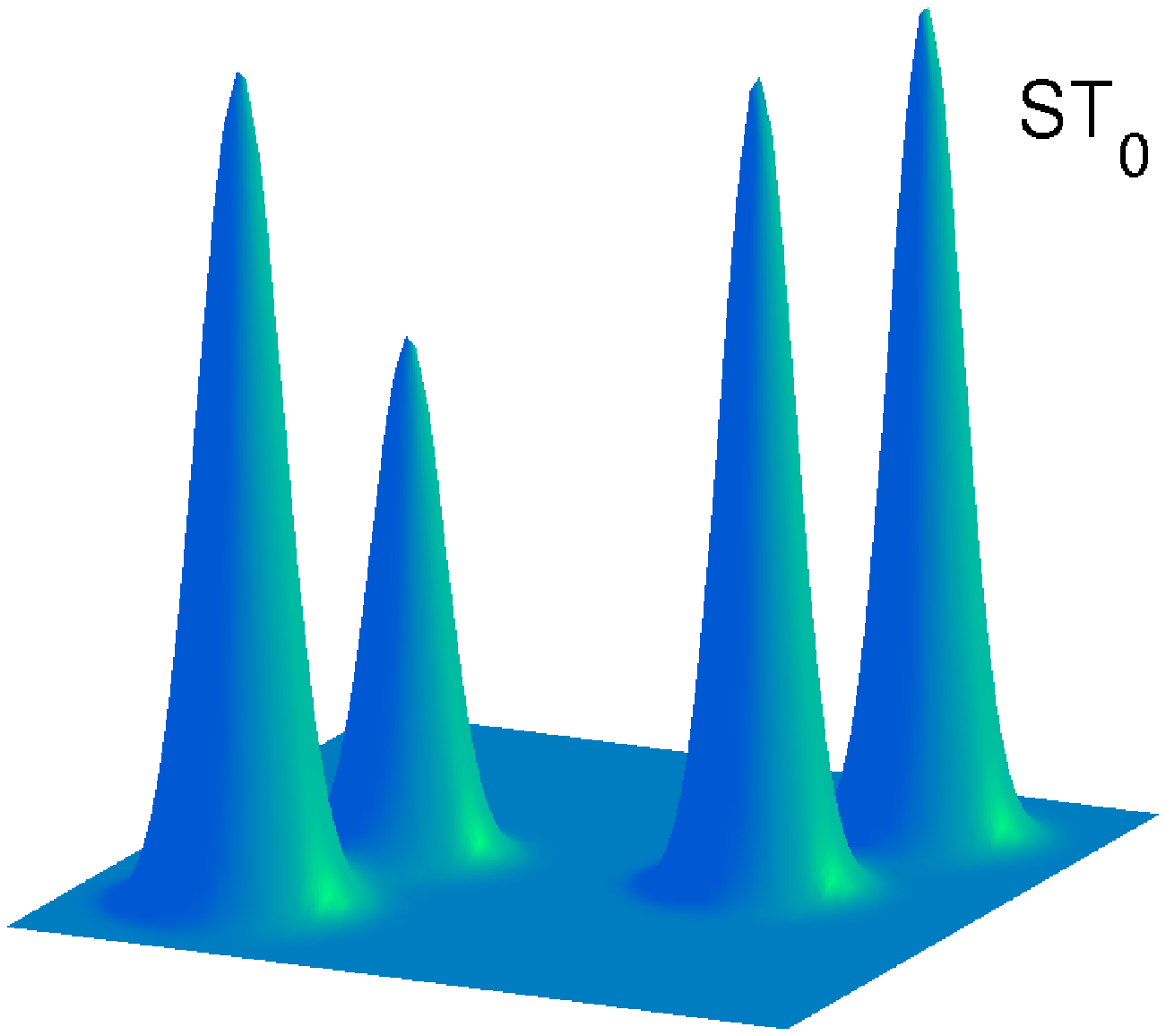}}
\subfigure{\includegraphics[width=.49\columnwidth]{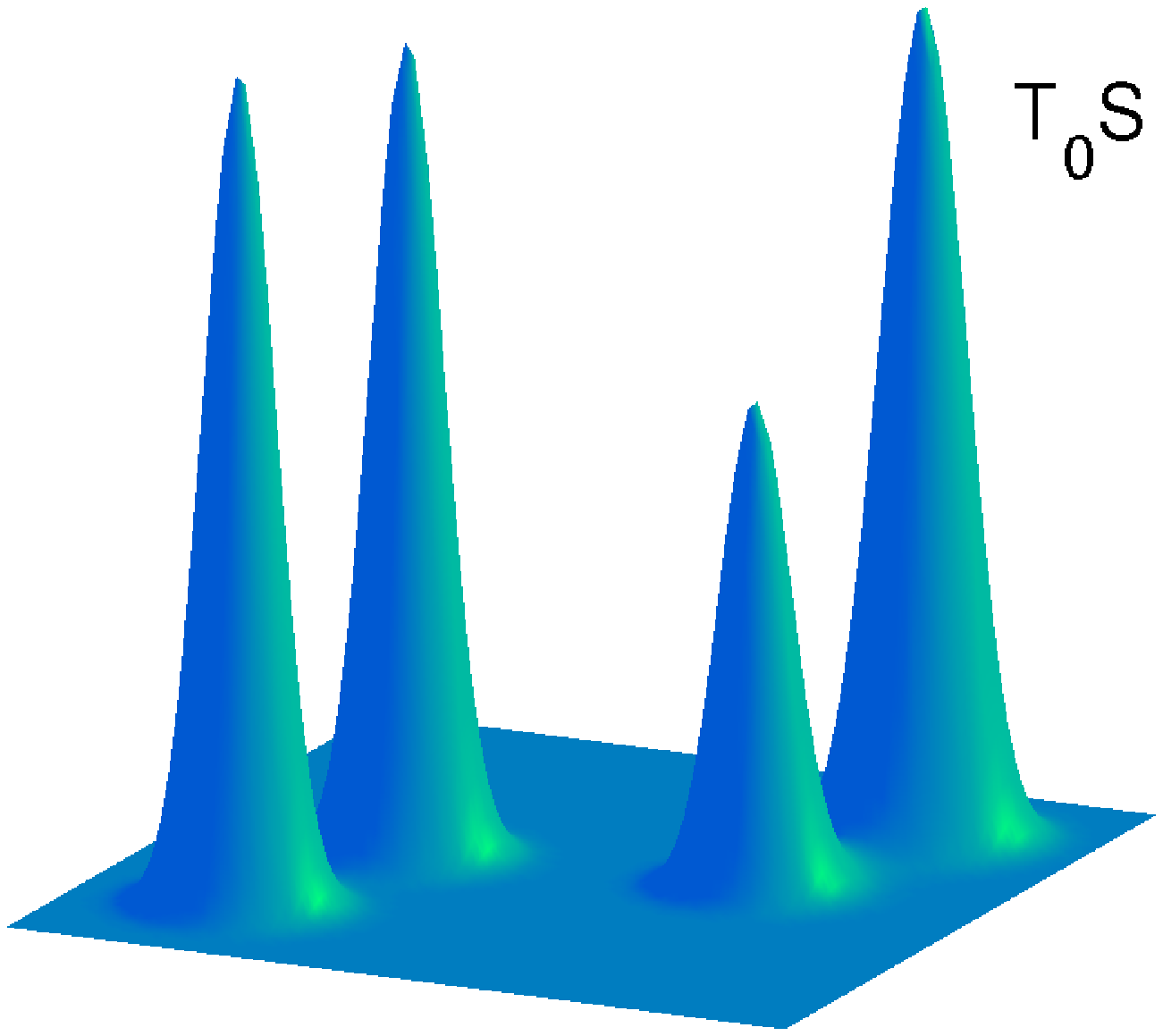}}
\subfigure{\includegraphics[width=.49\columnwidth]{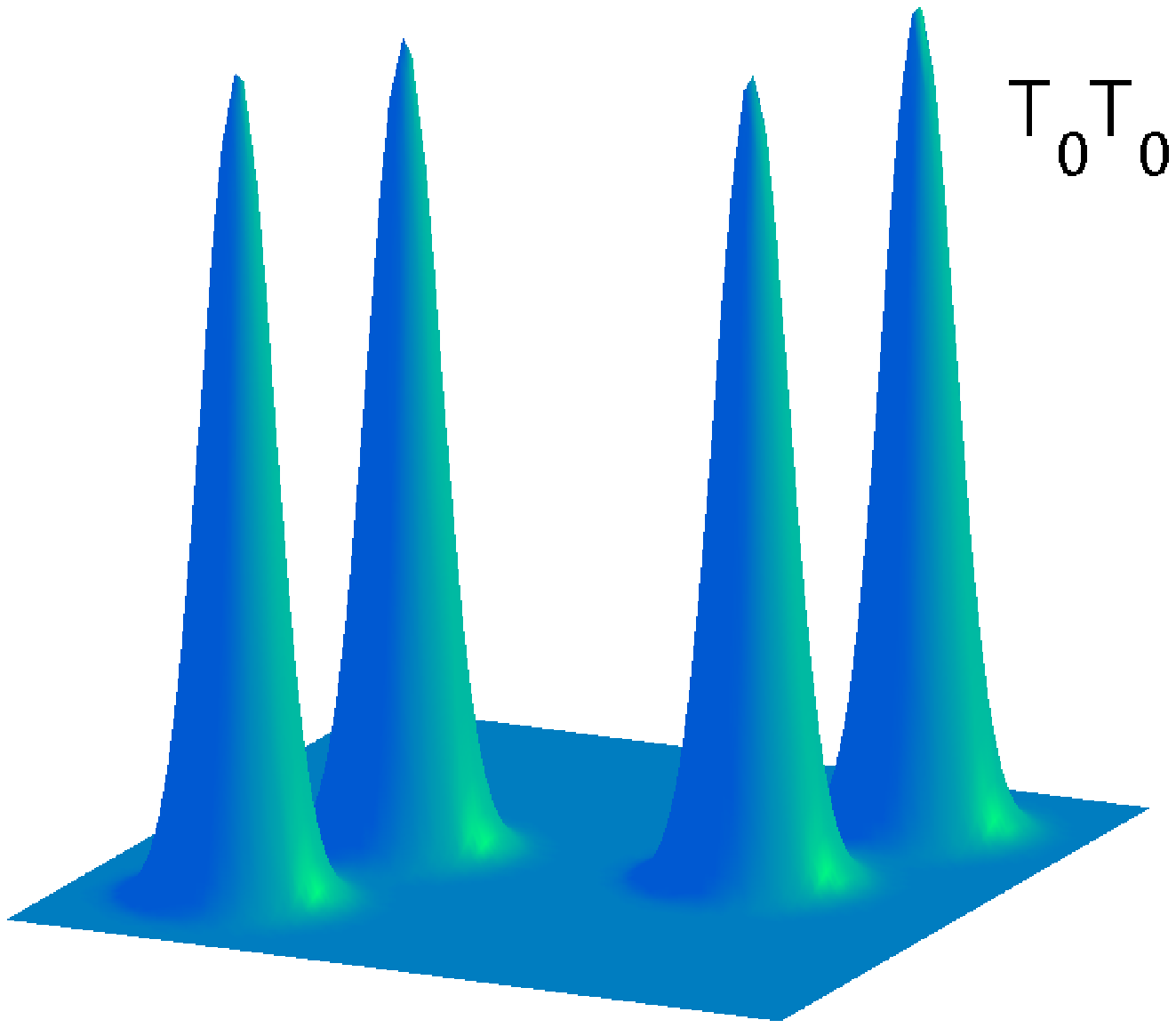}}
\caption{The charge density of the lowest eigenstates of the DQD system. The detunings are $\epsilon_A=\epsilon_B=4.3$ meV (dots $1$ and $4$ have low potential). The upper left plot shows the $|SS\rangle$ state,
upper right the $|ST_0\rangle$ state, lower left the $|T_0S\rangle$ state, and lower right the density of the triplet states (identical to the density of $|T_0T_0\rangle$).
}
\label{fig:densities}
\end{figure}

With non-zero detuning, one begins to see differences in the charge densities of the lowest eigenstates. Fig. \ref{fig:detuning} shows the effect of the sign of the detunings on the ground state $|SS\rangle$.
The charge densities of the lowest states (given by Lanczos) in the detuned case, $\epsilon_A=\epsilon_B=4.35$ meV  are shown in Fig. \ref{fig:densities}. 

There is a difference in the densities depending on which of the dots have the low potential, as can be seen in Fig. \ref{fig:detuning}.
In the upper right corner of Fig. \ref{fig:detuning} (and also in Fig. \ref{fig:densities}) the dots that are the furthest apart from each other have the lowest potential. This facilitates the localization of the singlet state
in these dots, as it reduces the Coulomb repulsion. In the cases when dots $2$ and $4$ or $1$ and $3$ are in the low potential, the singlet localizes only in the further away dots, $1$ and $4$.
In the case where dots $2$ and $3$ have the low detuning, there are four identical peaks, the singlets cannot localize in the neighboring dots due to the Coulomb repulsion.

The dipole-dipole entanglement effect relies on the differences of the singlet and triplet densities. Thus, to this end, the optimal detuning configuration should be as in Fig. \ref{fig:densities}, the dots
furthest away are detuned to low potential energy.
The densities in Fig. \ref{fig:densities} show that the singlets localize to the dots that have lower potential (dots $1$ and $4$). The fourth plot represents all the triplet states
and is identical also to the $|T_0T_0\rangle$ density. It shows four identical peaks, with exactly one electron in each dot. When the detuning is further increased, the singles localize fully to
the low lying dots, and at very high detunings, the triples start to localize as well. 

With high detuning ($\epsilon_A$ and $\epsilon_B$ above $5$ meV), 
$|SS\rangle$, $|ST_0\rangle$, and $|T_0S\rangle$ were still lowest in energy.
However, the triplet states were not the next three in this case. There are states lower in energy than the triplets, including other instances of the states $|SS\rangle$, $|ST_0\rangle$ and $|T_0S\rangle$ (i.e. states that are of the form $|X\rangle_A\otimes|Y\rangle_B$, where $|X\rangle$ and $|Y\rangle$ are $s=0$ or $s=1$
eigenstates of the two-electron $S^2$-operator).
\begin{figure}[!h]
\vspace{0.3cm}
\includegraphics[width=\columnwidth]{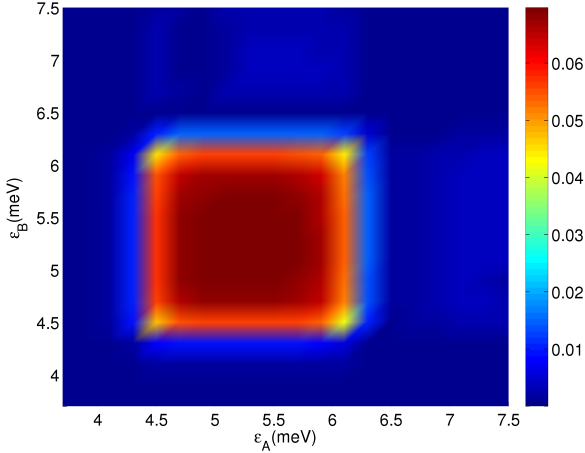}
\caption{(Color online) $|\Delta_E|$ as a function of the detunings $\epsilon_A$ and $\epsilon_B$. The values of $|\Delta_E|$ are shown in meV.
A high value of $|\Delta_E|$ means fast qubit-qubit coupling. The detunings are sampled at intervals of $0.2$ meV.}
\label{fig:delta}
\end{figure}
Next we consider states $|SS\rangle$, $|ST_0\rangle$, $|T_0S\rangle$, and $|T_0T_0\rangle$ that have the lowest energy and study the dependence of the coupling parameter $\Delta_E=E_{SS}+E_{T_0T_0}-E_{ST_0}-E_{T_0S}$ on the detunings $\epsilon_A$ and $\epsilon_B$ was studied.
A high value of $|\Delta_E|$ means fast qubit-qubit coupling, as seen in Eq. (\ref{eq:conc2}). 
In Fig. \ref{fig:delta}, $|\Delta_E|$ is shown as a function of the detunings.

The area of high $|\Delta_E|$ ($\epsilon_A,\epsilon_B$ between $4.3$ meV and $6.5$ meV) is roughly rectangular. Because
the transitions from $S(1,1)$ to $S(2,0)$ and from $T_0(1,1)$ to $T_0(2,0)$ happen quite abruptly with respect to increasing detuning.
In the low detuning region, the singlet and triplet states are both localized in two dots of the qubits and in the very high detuning case even the triplets localize
to dots $1$ and $4$. In the rectangular high $|\Delta_E|$-area, the singlets are in $(2,0)$-configuration and triplets in $(1,1)$ configuration.

Outside of the main rectangular area, there are also areas of smaller increase in $|\Delta_E|$. These are located at the sides of the rectangular peak,
and probably result from the fact that one of the qubits is in the detuning interval $4.3$ meV and $6.5$ meV. These side areas
were also present in the computations done by Nielsen et al, in addition to a large plateau in $|\Delta_E|$ when both detunings are high\cite{Nielsen}. This
plateau is not present in our results, possibly due to the fact that our qubit-qubit distance is quite large.

We then use our singlet-triplet qubit model, introduced in the previous section, to simulate the entanglement of the qubits $A$ and $B$ by the dipole-dipole interaction. We start from the ground state of the system, $|SS\rangle$ with zero detuning. The magnetic
field gradients in the qubits are then turned on, and the state of the system is let to evolve. When the qubit reach the $xy$-plane in the Bloch sphere, the magnetic field is turned off, and the detunings are turned on.
When the detunings have reached their maximum values, the system was is to evolve again. The evolution is computed according to Eq. (\ref{eq:prop}), and the concurrence according to Eq. (\ref{eq:conc}).

\begin{figure}[!h]
\vspace{0.3cm}
\includegraphics[width=\columnwidth]{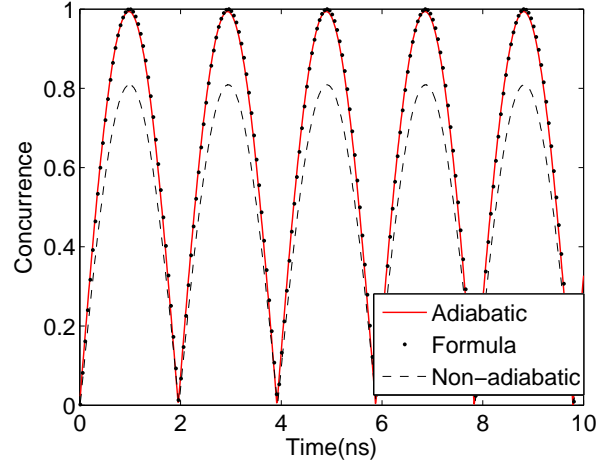}
\caption{(Color online) Entanglement of two singlet-triplet qubits measured by concurrence. The qubits were brought to the $xy$-plane of the Bloch sphere and let to evolve
for $10$ ns. The detunings were $\epsilon_A=\epsilon_B=4.3$ meV. The figure shows only the evolution in the $xy$-plane, as the concurrence was zero before this. Solid red
line shows the curve that was obtained with adiabatic increase of the detunings during $1$ ns. In the dashed curve, the detunings were turned on instantaneously. 
The curves are obtained by projecting the wave function onto the computational basis. The black markers show the concurrence given by the analytical formula Eq. (\ref{eq:conc2}).}
\label{fig:conc}
\end{figure}

The computed concurrences can be seen in Fig. \ref{fig:conc}. Here, the qubits are first brought to the $xy$-plane using magnetic field gradients of
$\Delta B_{z,A}=\Delta B_{z,B}=10$ mT. We study the effect of the speed by which the detunings are increased to the values $\epsilon_A=\epsilon_B=4.3$ meV. Cases of an instantaneous increase and an adiabatic increase during a time of $1$ ns
can be seen in Fig. \ref{fig:conc}.
The qubits are let to evolve in the $xy$ plane for a time of $10$ ns, and the concurrence is computed at each time step according to Eq. (\ref{eq:conc}) (the time step length was $\Delta t=1$ ps).
Fig. \ref{fig:conc} also shows the concurrence given by formula Eq. (\ref{eq:conc2}), where the energies are computed by Lanczos ($E_{T_0T_0}$ was obtained
as the eigenenergy of one of the triplet states).

In the adiabatic case, the concurrence oscillates between $0$ and $1$ as the state of the qubits is evolved in the $xy$-plane.
The obtained curve coincides almost completely with the one given by Eq. (\ref{eq:conc2}).
In the non-adiabatic case, the frequency of the concurrence oscillations, given by $\Delta_E$, is the same as in the adiabatic one, but
the amplitude is smaller, only about $0.81$. The reason for this is that when the detunings are increased instantaneously,
the wave function leaks from the computational basis. Indeed, the the probability for being in the computational basis, $\sum_{i,j=1}^2|M_{ij}|^2$, is only $0.82$
in this case (during the evolution in the $xy$-plane).

The forms of the computed concurrence curves are similar to the measurement by Shulman et al.\cite{shulman} apart from the decoherence effects
that have not yet been implemented in our model.
Fig. \ref{fig:conc} shows that
the formula in Eq. \ref{eq:conc2} indeed describes well the entanglement of qubits in our system if the detunings are increased gradually. It is however limited to the case of $xy$-rotations. In the more complex cases of time-dependent detunings and magnetic fields,
the concurrence can be computed by projecting the wave function onto the computational basis. In order to obtain the maximal degree of entanglement, the
detunings should be increased adiabatically to their maximal values.

It should be noted that it is not certain that the 24 first single particle states are enough to produce quantitatively accurate results in this two-DQD case when the detunings are very high. We have not
studied the convergence of, for example, the lowest four-particle energies by comparing the 24-basis results to ones obtained with a larger basis. Hence, the analysis in this section should be considered
only qualitatively valid. 
One could of course compute more states, but increasing the single particle basis size will result in significant increase in the four-particle basis used in the computation of dynamics.
The full dynamics computations were quite slow in the 24 state case already, and as the main topic of this article is the presentation of our method, quantitative analyses
are not included.

Instead of doing the matrix exponent for the full four-particle Hamiltonian,
one could first project it to some number of eigenstates of the Hamiltonian with zero detuning and magnetic field gradients. This small matrix can then be diagonalized exactly. This method was tested in the case of the 12 first four-particle
eigenstates (the number of single particle states in the Lanczos was 24), and it gave the same results as the full dynamics computations. By using this projection method,
one could increase the single particle basis size to obtain more accurate results without the expense in simulation lengths.

\section{Conclusions}

In summary, we presented a Lagrange mesh based scheme for studying many-particle states in lateral quantum dots. We introduced a Lagrange mesh based 
method for many-body systems and then proceeded to use the method in simulating singlet-triplet qubits. We introduced a model for simulating
the full quantum control of singlet-triplet DQD qubits, and showed that this model can be used to produce realistic dynamics of the qubit system. The singlet-triplet qubit model was
then used to study a system of two qubits. We computed and studied the lowest eigenstates of this system and also discussed the effect of electrostatic detuning
on the eigenstates. The entanglement of the two qubits via the dipole-dipole interaction was simulated.

The Lagrange mesh provides a very flexible method of dealing with complex confinement potentials in ED calculations. It allowed us to create a realistic first principles
model for studying the interplay and dynamics of two singlet-triplet qubits. This method could be used to study complex effects that are difficult to include in simpler models. In addition, our model could quite easily be further improved by the inclusion of the decoherence effects due to the
environment, like in\cite{jani1,jani2}.

\section*{Acknowledgements}

We thank Eero T\"ol\"o for his help with the analytic manipulation
of the interaction integrals. We acknowledge the support from Academy
of Finland through its Centers of Excellence Program (project no. 251748).

\section*{Appendix: Sinc mesh and Gauss quadrature}

In this appendix, we consider the validity of the Gauss quadrature approximation of the potential energy matrix elements,
\begin{equation}\label{eq:approx}
V_{\va'\va}\approx V(\va)\delta_{\va'\va},
\end{equation}
in the case of the sinc mesh. We mainly concentrate on the case of a parabolic potential, as this paper deals with locally parabolic quantum dots.
General results concerning other potentials are briefly discussed.

In principle, the potential matrix elements $V_{\va',\va}$ can be computed analytically,
\begin{equation}\label{eq:anal}
V_{\va'\va}=\int_{-\infty}^{\infty}\d \v{r} L_{\va'}(\v{r})V(\v{r})L_{\va}(\v{r}).
\end{equation}
The approximation (\ref{eq:approx}) can be considered valid if the results (i.e eigenstates and energies) obtained with it do not
deviate considerably from the ones obtained with the analytical formula.

Consider a one dimensional sinc mesh over the interval $(-\frac{L}{2},\frac{L}{2})$. The number of basis functions is $N$ and the grid
spacing is $h=\frac{L}{N}$. Unfortunately, the analytic integral for the matrix elements,
\begin{equation}\label{eq:anal2}
V_{a'a}=\frac{1}{h}\int_{-\infty}^{\infty}\d x\sinc\left[\frac{\pi}{h}(x-x_{a'})\right]V(x)\sinc\left[\frac{\pi}{h}(x-x_{a})\right],
\end{equation}
is divergent for example in the case of a parabolic potential, $V(x)=x^2$. In Eq. (\ref{eq:anal2}), $x_a=-\frac{L}{2}+ha$ and $x_{a'}=-\frac{L}{2}+ha'$.
The integration can, however, be done over some large interval $(-M,M)$, where
$M$ is a multiple of $h$, $M=\tilde{N}h$ and $\tilde{N}\geq N$. In this case, with the parabolic potential, the elements are given as
\begin{eqnarray}
\nonumber
&V_{a'a}&=x_a^2\delta_{a'a}+(-1)^{a'+a}\frac{2hM}{\pi^2}+\\
&(-1)^{a'+a}&\frac{(x_a+x_{a'})h}{\pi^2}\int_{-M-x_a}^{-M+x_a}\frac{\sin^2\left(\frac{\pi}{h}t\right)}{t}\d t \label{eq:diag},
\end{eqnarray}
where $\delta_{a'a}$ is the Kronecker delta.
The third term of (\ref{eq:diag}) can be approximated as
\begin{equation}\label{eq:second}
\begin{aligned}
h\left|\int_{-M-x_a}^{-M+x_a}\frac{\sin^2\left(\frac{\pi}{h}t\right)}{t}\d t\right| \\
\leq h \min_{u=a,a'}\left(\log\left|\frac{\frac{\tilde{N}}{2}+u}{\frac{\tilde{N}}{2}-u}\right|\right).
\end{aligned}
\end{equation}
This vanishes when $\tilde{N}\rightarrow\infty$ or $h\rightarrow0$. Unfortunately, the second term of (\ref{eq:diag}), 
$\frac{2hM}{\pi^2}=\frac{2L^2\tilde{N}}{\pi^2N^2}$ generally diverges, as $\tilde{N}$
approaches infinity (i.e. we approach the analytical formula (\ref{eq:anal2})). However, this also can converge to zero in the limit $\tilde{N}\rightarrow\infty$
if $h$ at the same time approaches zero. This is the case for example if $\tilde{N}=\text{ceiling}\left(\log N\right)N$. Thus, given small enough $h$, the potential energy matrix obtained analytically is arbitrarily close to the diagonal one obtained with the Gauss quadrature.

In the two dimensional case, the analytical integrals reduce to the same one dimensional integrals discussed above. However, due to computational limitations
$N$ cannot be very large in 2D when computing the two body interaction matrix elements ($N=17$ is the current maximum for the computation of the $V_{ijkl}$-elements because of the GPU memory limitations).
With realistic values of $L$, the off-diagonal terms of (\ref{eq:diag}) are non-negligible with $N=17$.

Consider for example the case of a parabolic well,
$V(\v{r})=\frac{1}{2}m^*\omega_0^2r^2$. The values of the constants are $m^*=0.067m_e$, $\hbar\omega_0=4$ meV, $L=200$ nm, $\tilde{L}=6L$ and $N=17$.
In atomic units, the absolute value of the off-diagonal elements is approximately 0.5 (the maximum being 0.501), while the maximum diagonal value is 7.184 (The maximum diagonal value
contains the maximum value of the potential, $\frac{1}{8}m^*\omega^2L^2$. Most of the diagonal values are much smaller than this). Thus, the diagonal Gauss
quadrature approximation does not seem to hold.

However, the eigenenergies computed with this analytically obtained potential matrix are very close to 
ones obtained with the approximation. For example, the relative differences between the 24 lowest eigenenergies are all below $10^{-5}$ (the energies from the approximation are also
very close to the exact solution of the problem, the relative differences being again less than $10^{-5}$). Similar accuracy holds for the eigenfunctions. Higher the energy the bigger the differences are,
but nevertheless the accuracy of the Gauss quadrature seems quite unexpected. Next, we try to find reason behind this phenomenon.

For simplicity, consider again the one dimensional case. Similar arguments apply to the 2D case. Let $\mathbf{u}$ be an eigenvector of the Hamiltonian
$\mathbf{H}$, where the potential matrix is computed with the Gauss quadrature approximation,
$\mathbf{H}\mathbf{u}=E\mathbf{u}$.
In order for the approximation to be valid, $\mathbf{u}$ should also be an approximate eigenvector of the Hamiltonian $\tilde{\mathbf{H}}$, where the potential
energy matrix is computed analytically, according to (\ref{eq:diag}). Here, we set the interval of integration $\tilde{L}$ to be large
enough that the term in (\ref{eq:second}) can be neglected, $\tilde{N}>>N$. Now, $\tilde{\mathbf{H}}$ can be written as
\begin{equation}\label{eq:delta}
\tilde{\mathbf{H}}=\mathbf{H}+\mathbf{\Delta},\Delta_{a'a}=(-1)^{a+a'}\delta,
\end{equation}
where $\delta=\frac{2L^2\tilde{N}}{\pi^2N^2}$.

The symmetric matrix $\mathbf{\Delta}$ is of rank $1$, its only non-zero eigenvalue being $\lambda_{\delta}=N\delta$.
The corresponding eigenvector is $\frac{1}{\sqrt{N}}\left[1~-1~ 1~...~(-1)^N\right]$. By projecting $\mathbf{u}$ onto the (mutually orthogonal) eigenvectors of $\mathbf{\Delta}$, one
can compute the product $\mathbf{\Delta u}$. As, apart from $\lambda_{\delta}$, all eigenvalues of $\mathbf{\Delta}$ are zero, it holds that
\begin{equation}
\tilde{\mathbf{H}}\mathbf{u}=E\mathbf{u}+\mathbf{\Delta}\mathbf{u}=E\mathbf{u}+\delta(\mathbf{v}^T\mathbf{u})\mathbf{v},
\end{equation}
where $\mathbf{v}=\left[1~-1~ 1~...~(-1)^N\right]$, and $\mathbf{v}^T\mathbf{u}=u_1-u_2+u_3-u_4+...+(-1)^Nu_N$.

We now reason why $\mathbf{v}^T\mathbf{u}$ should be small. 
By the Lagrange condition, $u_a=h\psi(x_a)$, where $\psi(x)=\sum_au_aL_a(x)$. As $\psi$ is a finite sum of differentiable functions, it also is 
a differentiable function. Thus, $\mathbf{u}$ is a discretization of a differentiable function.
Hence, the increments $u_a-u_{a+1}$ should be small and cancel each other out, as (with large enough $L$) $u_1\approx0$ and $u_N\approx0$.

This argument can be further justified by writing $u_a=u(x_a)$, and
noticing that $\mathbf{v}^T\mathbf{u}$ is related to the integral of the derivative of $u$.
\begin{eqnarray*}
\mathbf{v}^T\mathbf{u}&=&\frac{u(x_1)-u(x_1+h)}{h}h+\frac{u(x_3)-u(x_3+h)}{h}h+...\\
&\approx&-\frac{1}{2}\sum_{j}u'(x_{2j+1})2h \\
&\approx&-\frac{1}{2}­\int_{-L/2}^{L/2}u'(t)dt=\frac{1}{2}\left[u(-L/2)-u(L/2)\right].
\end{eqnarray*}
If the system area $L$ is large enough, $\psi(\pm L/2)\approx0$. Furthermore, as $u(x)=h\psi(x)$, $u(-L/2)-u(L/2)$ can be considered to be very close to zero.

Now, we can approximate the error term $\mathbf{\Delta}\mathbf{u}$
using an error formula for Riemann-sums. As we have approximated the ratio of differences, $\frac{\Delta u}{\Delta x}$ of $u(x)$, to be the derivative of $u$, some error
arises from this approximation as well. From the Taylor expansion: $\frac{\Delta u}{\Delta x}h=u'(x)+O(h^3)=h\psi'(x)+O(h^3)$, where we have used the fact that $u(x)=h\psi(x)$. 
The error for the Riemann sum formula is in this case smaller than $\frac{DL^3}{N^2}$, where $D=\max_{x}|\psi'(x)|$. The elements of the 'error' vector $\mathbf{\Delta}\mathbf{u}=\delta(\mathbf{v}^T\mathbf{u})\mathbf{v}$ thus obey
\begin{equation}\label{eq:error}
|(\Delta u)_a|\leq\frac{\delta DL^3}{N^2}+\delta O(h^3)=\frac{DL^5\tilde{N}}{\pi^2N^4}+O(N^{-5}).
\end{equation}

We have shown that the elements of $\mathbf{\Delta}\mathbf{u}$ behave as $\sim\frac{\tilde{N}}{N^4}$. They become small even with relatively small values of $N$. Eq. (\ref{eq:error}) also explains why higher eigenenergies tend to differ more between
the Gauss approximation and the analytical formulas. Higher eigenstates oscillate more and thus have larger values of $|\psi'(x)|$.

Same kind of analysis can be done to potentials $V(x)=x^k,k\in\mathbb{N}$ or $V(\v{r})=r^k,k\in\mathbb{N}$, and the results are similar; the analytical formula
for the potential energy matrix elements gives approximately the same eigenstates as the Gauss quadrature. The matrix $\mathbf{\Delta}$ is of the same alternating form in all of these cases
as in Eq. (\ref{eq:delta}), and thus the same convergence arguments apply here. In principle, the result can then be extended to any potential that can be written
as a power series.

In conclusion, the analytic formula for the potential matrix elements of a quadratic well can be made to converge to the Gauss quadrature approximation when the grid spacing goes to zero. In addition,
the accuracy of the Gauss quadrature eigenenergies and states remains good even when the approximation does not hold for the individual matrix elements. Similar results
apply for other potentials as well, for example $r^k,k\in\mathbb{N}$.

\bibliography{lagrange}

\begin{thebibliography}{10}%
\makeatletter
\providecommand \@ifxundefined [1]{%
 \ifx #1\undefined \expandafter \@firstoftwo
 \else \expandafter \@secondoftwo
\fi
}%
\providecommand \@ifnum [1]{%
 \ifnum #1\expandafter \@firstoftwo
 \else \expandafter \@secondoftwo
\fi
}%
\providecommand \enquote [1]{``#1''}%
\providecommand \bibnamefont  [1]{#1}%
\providecommand \bibfnamefont [1]{#1}%
\providecommand \citenamefont [1]{#1}%
\providecommand\href[0]{\@sanitize\@href}%
\providecommand\@href[1]{\endgroup\@@startlink{#1}\endgroup\@@href}%
\providecommand\@@href[1]{#1\@@endlink}%
\providecommand \@sanitize [0]{\begingroup\catcode`\&12\catcode`\#12\relax}%
\@ifxundefined \pdfoutput {\@firstoftwo}{%
 \@ifnum{\z@=\pdfoutput}{\@firstoftwo}{\@secondoftwo}%
}{%
 \providecommand\@@startlink[1]{\leavevmode\special{html:<a href="#1">}}%
 \providecommand\@@endlink[0]{\special{html:</a>}}%
}{%
 \providecommand\@@startlink[1]{%
  \leavevmode
  \pdfstartlink
   attr{/Border[0 0 1 ]/H/I/C[0 1 1]}%
   user{/Subtype/Link/A<</Type/Action/S/URI/URI(#1)>>}%
  \relax
 }%
 \providecommand\@@endlink[0]{\pdfendlink}%
}%
\providecommand \url  [0]{\begingroup\@sanitize \@url }%
\providecommand \@url [1]{\endgroup\@href {#1}{\urlprefix}}%
\providecommand \urlprefix [0]{URL }%
\providecommand \Eprint[0]{\href }%
\@ifxundefined \urlstyle {%
  \providecommand \doi [1]{doi:\discretionary{}{}{}#1}%
}{%
  \providecommand \doi [0]{doi:\discretionary{}{}{}\begingroup
  \urlstyle{rm}\Url }%
}%
\providecommand \doibase [0]{http://dx.doi.org/}%
\providecommand \Doi[1]{\href{\doibase#1}}%
\providecommand \bibAnnote [3]{%
  \BibitemShut{#1}%
  \begin{quotation}\noindent
    \textsc{Key:}\ #2\\\textsc{Annotation:}\ #3%
  \end{quotation}%
}%
\providecommand \bibAnnoteFile [2]{%
  \IfFileExists{#2}{\bibAnnote {#1} {#2} {\input{#2}}}{}%
}%
\providecommand \typeout [0]{\immediate \write \m@ne }%
\providecommand \selectlanguage [0]{\@gobble}%
\providecommand \bibinfo [0]{\@secondoftwo}%
\providecommand \bibfield [0]{\@secondoftwo}%
\providecommand \translation [1]{[#1]}%
\providecommand \BibitemOpen[0]{}%
\providecommand \bibitemStop [0]{}%
\providecommand \bibitemNoStop [0]{.\EOS\space}%
\providecommand \EOS [0]{\spacefactor3000\relax}%
\providecommand \BibitemShut [1]{\csname bibitem#1\endcsname}%
\bibitem{Ashoori96}%
  \BibitemOpen
  \bibfield{author}{%
  \bibinfo {author} {\bibfnamefont{R.~C.}\ \bibnamefont{Ashoori}},\ }%
  \bibfield{journal}{%
  \bibinfo {journal} {Nature}\ }%
  \textbf{\bibinfo {volume} {379}},\ \bibinfo {pages} {413} (\bibinfo {year}
  {1996})%
  \bibAnnoteFile{NoStop}{Ashoori96}%
\bibitem{Reimann02}%
  \BibitemOpen
  \bibfield{author}{%
  \bibinfo {author} {\bibfnamefont{S.~M.}\ \bibnamefont{Reimann}}\ and\
  \bibinfo {author} {\bibfnamefont{M.}~\bibnamefont{Manninen}},\ }%
  \bibfield{journal}{%
  \Doi{10.1103/RevModPhys.74.1283}{\bibinfo {journal} {Rev. Mod. Phys.}}\ }%
  \textbf{\bibinfo {volume} {74}},\ \bibinfo {pages} {1283} (\bibinfo {year}
  {2002})%
  \bibAnnoteFile{NoStop}{Reimann02}%
\bibitem{Saarikoski_RMP}%
  \BibitemOpen
  \bibfield{author}{%
  \bibinfo {author} {\bibfnamefont{H.}~\bibnamefont{Saarikoski}}, \bibinfo
  {author} {\bibfnamefont{S.~M.}\ \bibnamefont{Reimann}}, \bibinfo {author}
  {\bibfnamefont{A.}~\bibnamefont{Harju}},\ and\ \bibinfo {author}
  {\bibfnamefont{M.}~\bibnamefont{Manninen}},\ }%
  \bibfield{journal}{%
  \Doi{10.1103/RevModPhys.82.2785}{\bibinfo {journal} {Rev. Mod. Phys.}}\ }%
  \textbf{\bibinfo {volume} {82}},\ \bibinfo {pages} {2785} (\bibinfo {year}
  {2010})%
  \bibAnnoteFile{NoStop}{Saarikoski_RMP}%
\bibitem{Loss98}%
  \BibitemOpen
  \bibfield{author}{%
  \bibinfo {author} {\bibfnamefont{D.}~\bibnamefont{Loss}}\ and\ \bibinfo
  {author} {\bibfnamefont{D.~P.}\ \bibnamefont{DiVincenzo}},\ }%
  \bibfield{journal}{%
  \Doi{10.1103/PhysRevA.57.120}{\bibinfo {journal} {Phys. Rev. A}}\ }%
  \textbf{\bibinfo {volume} {57}},\ \bibinfo {pages} {120} (\bibinfo {year}
  {1998})%
  \bibAnnoteFile{NoStop}{Loss98}%
\bibitem{Hanson07}%
  \BibitemOpen
  \bibfield{author}{%
  \bibinfo {author} {\bibfnamefont{R.}~\bibnamefont{Hanson}}, \bibinfo {author}
  {\bibfnamefont{L.~P.}\ \bibnamefont{Kouwenhoven}}, \bibinfo {author}
  {\bibfnamefont{J.~R.}\ \bibnamefont{Petta}}, \bibinfo {author}
  {\bibfnamefont{S.}~\bibnamefont{Tarucha}},\ and\ \bibinfo {author}
  {\bibfnamefont{L.~M.~K.}\ \bibnamefont{Vandersypen}},\ }%
  \bibfield{journal}{%
  \Doi{10.1103/RevModPhys.79.1217}{\bibinfo {journal} {Rev. Mod. Phys.}}\ }%
  \textbf{\bibinfo {volume} {79}},\ \bibinfo {pages} {1217} (\bibinfo {year}
  {2007})%
  \bibAnnoteFile{NoStop}{Hanson07}%
\bibitem{levy}%
  \BibitemOpen
  \bibfield{author}{%
  \bibinfo {author} {\bibfnamefont{J.}~\bibnamefont{Levy}},\ }%
  \bibfield{journal}{%
  \bibinfo {journal} {Physical review letters}\ }%
  \textbf{\bibinfo {volume} {89}},\ \bibinfo {pages} {147902} (\bibinfo {year}
  {2002})%
  \bibAnnoteFile{NoStop}{levy}%
\bibitem{taylor}%
  \BibitemOpen
  \bibfield{author}{%
  \bibinfo {author} {\bibfnamefont{J.}~\bibnamefont{Taylor}}, \bibinfo {author}
  {\bibfnamefont{H.}~\bibnamefont{Engel}}, \bibinfo {author}
  {\bibfnamefont{W.}~\bibnamefont{D{\"u}r}}, \bibinfo {author}
  {\bibfnamefont{A.}~\bibnamefont{Yacoby}}, \bibinfo {author}
  {\bibfnamefont{C.}~\bibnamefont{Marcus}}, \bibinfo {author}
  {\bibfnamefont{P.}~\bibnamefont{Zoller}},\ and\ \bibinfo {author}
  {\bibfnamefont{M.}~\bibnamefont{Lukin}},\ }%
  \bibfield{journal}{%
  \bibinfo {journal} {Nat. Phys.}\ }%
  \textbf{\bibinfo {volume} {1}},\ \bibinfo {pages} {177} (\bibinfo {year}
  {2005})%
  \bibAnnoteFile{NoStop}{taylor}%
\bibitem{Pettaa}%
  \BibitemOpen
  \bibfield{author}{%
  \bibinfo {author} {\bibfnamefont{J.~R.}\ \bibnamefont{Petta}}, \bibinfo
  {author} {\bibfnamefont{H.}~\bibnamefont{Lu}},\ and\ \bibinfo {author}
  {\bibfnamefont{A.~C.}\ \bibnamefont{Gossard}},\ }%
  \bibfield{journal}{%
  \Doi{10.1126/science.1183628}{\bibinfo {journal} {Science}}\ }%
  \textbf{\bibinfo {volume} {327}},\ \bibinfo {pages} {669} (\bibinfo {year}
  {2010})%
  \bibAnnoteFile{NoStop}{Pettaa}%
\bibitem{foletti}%
  \BibitemOpen
  \bibfield{author}{%
  \bibinfo {author} {\bibfnamefont{S.}~\bibnamefont{Foletti}}, \bibinfo
  {author} {\bibfnamefont{H.}~\bibnamefont{Bluhm}}, \bibinfo {author}
  {\bibfnamefont{D.}~\bibnamefont{Mahalu}}, \bibinfo {author}
  {\bibfnamefont{V.}~\bibnamefont{Umansky}},\ and\ \bibinfo {author}
  {\bibfnamefont{A.}~\bibnamefont{Yacoby}},\ }%
  \bibfield{journal}{%
  \bibinfo {journal} {Nat. Phys.}\ }%
  \textbf{\bibinfo {volume} {5}},\ \bibinfo {pages} {903} (\bibinfo {year}
  {2009})%
  \bibAnnoteFile{NoStop}{foletti}%
\bibitem{shulman}%
  \BibitemOpen
  \bibfield{author}{%
  \bibinfo {author} {\bibfnamefont{M.}~\bibnamefont{Shulman}}, \bibinfo
  {author} {\bibfnamefont{O.}~\bibnamefont{Dial}}, \bibinfo {author}
  {\bibfnamefont{S.}~\bibnamefont{Harvey}}, \bibinfo {author}
  {\bibfnamefont{H.}~\bibnamefont{Bluhm}}, \bibinfo {author}
  {\bibfnamefont{V.}~\bibnamefont{Umansky}},\ and\ \bibinfo {author}
  {\bibfnamefont{A.}~\bibnamefont{Yacoby}},\ }%
  \bibfield{journal}{%
  \bibinfo {journal} {Science}\ }%
  \textbf{\bibinfo {volume} {336}},\ \bibinfo {pages} {202} (\bibinfo {year}
  {2012})%
  \bibAnnoteFile{NoStop}{shulman}%
\bibitem{ent}%
  \BibitemOpen
  \bibfield{author}{%
  \bibinfo {author} {\bibfnamefont{R.}~\bibnamefont{Horodecki}}, \bibinfo
  {author} {\bibfnamefont{P.}~\bibnamefont{Horodecki}}, \bibinfo {author}
  {\bibfnamefont{M.}~\bibnamefont{Horodecki}},\ and\ \bibinfo {author}
  {\bibfnamefont{K.}~\bibnamefont{Horodecki}},\ }%
  \bibfield{journal}{%
  \bibinfo {journal} {Rev. Mod. Phys.}\ }%
  \textbf{\bibinfo {volume} {81}},\ \bibinfo {pages} {865} (\bibinfo {year}
  {2009})%
  \bibAnnoteFile{NoStop}{ent}%
\bibitem{li}%
  \BibitemOpen
  \bibfield{author}{%
  \bibinfo {author} {\bibfnamefont{R.}~\bibnamefont{Li}}, \bibinfo {author}
  {\bibfnamefont{X.}~\bibnamefont{Hu}},\ and\ \bibinfo {author}
  {\bibfnamefont{J.~Q.}\ \bibnamefont{You}},\ }%
  \bibfield{journal}{%
  \bibinfo {journal} {Phys. Rev. B}\ }%
  \textbf{\bibinfo {volume} {86}},\ \bibinfo {pages} {205306} (\bibinfo {year}
  {2012})%
  \bibAnnoteFile{NoStop}{li}%
\bibitem{hanson}%
  \BibitemOpen
  \bibfield{author}{%
  \bibinfo {author} {\bibfnamefont{R.}~\bibnamefont{Hanson}}\ and\ \bibinfo
  {author} {\bibfnamefont{G.}~\bibnamefont{Burkard}},\ }%
  \bibfield{journal}{%
  \bibinfo {journal} {Phys. Rev. Lett.}\ }%
  \textbf{\bibinfo {volume} {98}},\ \bibinfo {pages} {050502} (\bibinfo {year}
  {2007})%
  \bibAnnoteFile{NoStop}{hanson}%
\bibitem{stepa}%
  \BibitemOpen
  \bibfield{author}{%
  \bibinfo {author} {\bibfnamefont{D.}~\bibnamefont{Stepanenko}}\ and\ \bibinfo
  {author} {\bibfnamefont{G.}~\bibnamefont{Burkard}},\ }%
  \bibfield{journal}{%
  \bibinfo {journal} {Phys. Rev. B}\ }%
  \textbf{\bibinfo {volume} {75}},\ \bibinfo {pages} {085324} (\bibinfo {year}
  {2007})%
  \bibAnnoteFile{NoStop}{stepa}%
\bibitem{Nielsen}%
  \BibitemOpen
  \bibfield{author}{%
  \bibinfo {author} {\bibfnamefont{E.}~\bibnamefont{Nielsen}}, \bibinfo
  {author} {\bibfnamefont{R.~P.}\ \bibnamefont{Muller}},\ and\ \bibinfo
  {author} {\bibfnamefont{M.~S.}\ \bibnamefont{Carroll}},\ }%
  \bibfield{journal}{%
  \bibinfo {journal} {Phys. Rev. B}\ }%
  \textbf{\bibinfo {volume} {85}},\ \bibinfo {pages} {035319} (\bibinfo {year}
  {2012})%
  \bibAnnoteFile{NoStop}{Nielsen}%
\bibitem{VMC}%
  \BibitemOpen
  \bibfield{author}{%
  \bibinfo {author} {\bibfnamefont{A.}~\bibnamefont{Harju}},\ }%
  \bibfield{journal}{%
  \bibinfo {journal} {J. Low Temp. Phys.}\ }%
  \textbf{\bibinfo {volume} {140}},\ \bibinfo {pages} {181} (\bibinfo {year}
  {2005})%
  \bibAnnoteFile{NoStop}{VMC}%
\bibitem{Henri}%
  \BibitemOpen
  \bibfield{author}{%
  \bibinfo {author} {\bibfnamefont{H.}~\bibnamefont{Saarikoski}}\ and\ \bibinfo
  {author} {\bibfnamefont{A.}~\bibnamefont{Harju}},\ }%
  \bibfield{journal}{%
  \Doi{10.1103/PhysRevLett.94.246803}{\bibinfo {journal} {Phys. Rev. Lett.}}\
  }%
  \textbf{\bibinfo {volume} {94}},\ \bibinfo {pages} {246803} (\bibinfo {month}
  {Jun}\ \bibinfo {year} {2005})%
  \bibAnnoteFile{NoStop}{Henri}%
\bibitem{Baye86}%
  \BibitemOpen
  \bibfield{author}{%
  \bibinfo {author} {\bibfnamefont{D.}~\bibnamefont{Baye}}\ and\ \bibinfo
  {author} {\bibfnamefont{P.~H.}\ \bibnamefont{Heenen}},\ }%
  \bibfield{journal}{%
  \bibinfo {journal} {J. Phys. A}\ }%
  \textbf{\bibinfo {volume} {19}},\ \bibinfo {pages} {2041} (\bibinfo {year}
  {1986})%
  \bibAnnoteFile{NoStop}{Baye86}%
\bibitem{Baye06}%
  \BibitemOpen
  \bibfield{author}{%
  \bibinfo {author} {\bibfnamefont{D.}~\bibnamefont{Baye}},\ }%
  \bibfield{journal}{%
  \bibinfo {journal} {Phys. Stat. Sol. (B)}\ }%
  \textbf{\bibinfo {volume} {243}},\ \bibinfo {pages} {1095} (\bibinfo {year}
  {2006})%
  \bibAnnoteFile{NoStop}{Baye06}%
\bibitem{Varga04}%
  \BibitemOpen
  \bibfield{author}{%
  \bibinfo {author} {\bibfnamefont{K.}~\bibnamefont{Varga}}, \bibinfo {author}
  {\bibfnamefont{Z.}~\bibnamefont{Zhang}},\ and\ \bibinfo {author}
  {\bibfnamefont{S.~T.}\ \bibnamefont{Pantelides}},\ }%
  \bibfield{journal}{%
  \Doi{10.1103/PhysRevLetters93.176403}{\bibinfo {journal} {Phys. Rev. Lett.}}\
  }%
  \textbf{\bibinfo {volume} {93}},\ \bibinfo {pages} {176403} (\bibinfo {year}
  {2004})%
  \bibAnnoteFile{NoStop}{Varga04}%
\bibitem{CUDA}%
  \BibitemOpen
  \bibfield{author}{%
  \bibinfo {author} {\bibfnamefont{J.}~\bibnamefont{Nickolls}}, \bibinfo
  {author} {\bibfnamefont{I.}~\bibnamefont{Buck}}, \bibinfo {author}
  {\bibfnamefont{M.}~\bibnamefont{Garland}},\ and\ \bibinfo {author}
  {\bibfnamefont{K.}~\bibnamefont{Skadron}},\ }%
  \bibfield{journal}{%
  \Doi{http://doi.acm.org/10.1145/1365490.1365500}{\bibinfo {journal} {Queue}}\
  }%
  \textbf{\bibinfo {volume} {6}},\ \bibinfo {pages} {40} (\bibinfo {year}
  {2008})%
  \bibAnnoteFile{NoStop}{CUDA}%
\bibitem{jani1}%
  \BibitemOpen
  \bibfield{author}{%
  \bibinfo {author} {\bibfnamefont{J.}~\bibnamefont{S\"arkk\"a}}\ and\ \bibinfo
  {author} {\bibfnamefont{A.}~\bibnamefont{Harju}},\ }%
  \bibfield{journal}{%
  \bibinfo {journal} {Phys. Rev. B}\ }%
  \textbf{\bibinfo {volume} {77}},\ \bibinfo {pages} {245315} (\bibinfo {year}
  {2008})%
  \bibAnnoteFile{NoStop}{jani1}%
\bibitem{jani2}%
  \BibitemOpen
  \bibfield{author}{%
  \bibinfo {author} {\bibfnamefont{J.}~\bibnamefont{S\"arkk\"a}}\ and\ \bibinfo
  {author} {\bibfnamefont{A.}~\bibnamefont{Harju}},\ }%
  \bibfield{journal}{%
  \bibinfo {journal} {New J. Phys.}\ }%
  \textbf{\bibinfo {volume} {13}},\ \bibinfo {pages} {043010} (\bibinfo {year}
  {2011})%
  \bibAnnoteFile{NoStop}{jani2}%
\end{thebibliography}%

\end{document}